\documentclass[a4paper, 11pt]{article}
\usepackage{epsfig,amsmath,amsfonts}

\usepackage{bm}
\usepackage{bbm}
\usepackage{amssymb}
\usepackage{mathrsfs}
\usepackage{mathtools}
\usepackage{simplewick}
\usepackage{hyperref}

\DeclarePairedDelimiter{\ceil}{\lceil}{\rceil}
\DeclarePairedDelimiter{\floor}{\lfloor}{\rfloor}
\DeclarePairedDelimiter{\average}{\langle}{\rangle}
\newcommand{\h}[1]{h\mathopen{}\left(#1\right)\mathclose{}}
\newcommand{\hh}[2]{h\mathopen{}\left(#1+\tfrac 12 #2\right)\mathclose{}}
\newcommand{\hhh}[3]{h^{#3}\mathopen{}\left(#1+\tfrac 12 #2\right)\mathclose{}}
\newcommand{\hti}[1]{\tilde h\mathopen{}\left(#1\right)\mathclose{}}
\newcommand{\hhti}[2]{\tilde h\mathopen{}\left(#1+\tfrac 12 #2\right)\mathclose{}}

\newcommand{\ket}[1]{| #1\rangle}

\newcommand{\TorPDF}[2]{\texorpdfstring{$#1$}{#2}}

\newcommand{\diffp}[2]{\frac{\partial #1}{\partial #2}}
\newcommand{\ud}{\text{d}} %Better that DeclareMathOperator, since it doens't introduce too much whitespace after the upright d.
\DeclareMathOperator{\hs}{hs}
\DeclareMathOperator{\shs}{shs}
\DeclareMathOperator{\bpartial}{\bar{\partial}}

\DeclareMathSymbol{\negative}{\mathord}{operators}{"2D} %For ``negative'' sign, which is a small minus with less space.

\newcommand{\smallcirc}[1]{\scalebox{#1}{$\circ$}}
\DeclareMathOperator*{\sumCircles}{%
\mathchoice%
  {\ooalign{\phantom{$\displaystyle\sum$}\cr\hidewidth\raisebox{1.3\height}{$\mkern22mu\smallcirc{0.7}$}\hidewidth\cr%
                                  \raisebox{-0.75\height}{$\mkern22mu\smallcirc{0.7}$}\cr
                                  \hidewidth$\displaystyle\sum$}}
  {\ooalign{$\textstyle\sum$\cr%
                                \hidewidth\raisebox{1.9\height}{$\mkern16mu\smallcirc{0.4}$}\hidewidth\cr
                                \hidewidth\raisebox{-.3\height}{$\mkern16mu\smallcirc{0.4}$}\hidewidth\cr}}
  {\ooalign{\raisebox{0\height}{\scalebox{.6}{$\scriptstyle\sum$}}\cr%
                                \hidewidth\raisebox{1.6\height}{$\mkern7.5mu\smallcirc{0.2}$}\hidewidth\cr
                                \hidewidth\raisebox{-0.2\height}{$\mkern7.5mu\smallcirc{0.2}$}\hidewidth\cr}}
  {\ooalign{\raisebox{.2\height}{\scalebox{.6}{$\scriptstyle\sum$}}\cr%
                                \hidewidth\raisebox{2.2\height}{$\mkern7.5mu\smallcirc{0.2}$}\hidewidth\cr
                                \hidewidth\raisebox{0.4\height}{$\mkern7.5mu\smallcirc{0.2}$}\hidewidth\cr}}
}

\newcommand{\be}{\begin{equation}}
\newcommand{\ee}{\end{equation}}

\newcommand{\half}{\frac12}
\newcommand{\mhalf}{\mbox{\small $\frac12$}}

\newcommand{\Wcal}{\mathcal{W}}

\newcommand{\Ncal}{\mathcal{N}}

\newcommand{\AdS}{\mathrm{AdS}}

\newcommand{\Srm}{\mathrm{S}}

\newcommand{\p}{\partial}
\newcommand{\pbar}{\bar{\partial}}
\newcommand{\bt}{\tilde{b}}
\newcommand{\ct}{\tilde{c}}
\newcommand{\betat}{\tilde{\beta}}
\newcommand{\gammat}{\tilde{\gamma}}
\newcommand{\phit}{\tilde{\phi}}
\newcommand{\psit}{\tilde{\psi}}
\newcommand{\diff}{{\mathrm d}}
\newcommand{\comment}[1]{}
\newcommand{\zbar}{\bar{z}}

\newcommand{\hb}{\bar{h}}

\numberwithin{equation}{section}

\begin{document}

\begin{titlepage}

 \renewcommand{\thefootnote}{\fnsymbol{footnote}}

 \begin{center}
\mbox{}
\vspace{2cm}

\noindent{\Large \textbf{Three-Point Functions in $\mathcal N=2$ Higher-Spin Holography}}\\
\vspace{2cm}

\noindent{ Heidar Moradi and Konstantinos Zoubos\\
\vspace{2mm}
			\texttt{moradi@nbi.dk, kzoubos@nbi.dk}}
\bigskip

 \vskip .6 truecm

\centerline{\it Niels Bohr Institute
} \centerline{\it
Blegdamsvej 17, DK-2100 Copenhagen \O} 
\centerline{\it
Denmark}
 \vskip .4 truecm

 \end{center}

 \vspace{30mm}

\begin{abstract}

The $\mathbb CP^N$ Kazama-Suzuki models 
with the non-linear chiral algebra $\mathcal{SW}_\infty[\lambda]$ have been conjectured to be  dual to the fully 
supersymmetric Prokushkin-Vasiliev 
theory of higher-spin gauge fields coupled to two massive $\mathcal N=2$ multiplets on $\AdS_3$. We perform a non-trivial
check of this duality by computing three-point functions containing one higher-spin gauge field for arbitrary spin 
$s$ and deformation parameter $\lambda$ from the bulk theory, and from the boundary using a free ghost system based 
on the linear $sw_\infty[\lambda]$ algebra. We find an exact match between the two computations. 
In the 't Hooft limit, the three-point functions only depend on the wedge subalgebra $\shs[\lambda]$ and the 
results are equivalent for any theory with such a subalgebra. In the process we also find the emergence 
of $\mathcal N=2$ superconformal symmetry near the $\AdS_3$ boundary by computing holographic OPE's, 
consistently with a recent analysis of asymptotic symmetries of higher-spin supergravity.

\end{abstract}
\vfill
\vskip 0.5 truecm

\setcounter{footnote}{0}
\renewcommand{\thefootnote}{\arabic{footnote}}
\end{titlepage}

\newpage

\tableofcontents

\section{Introduction}

 Since its inception, the AdS/CFT correspondence \cite{Maldacena:1997re} has been 
one of the major research directions within the high energy theory community and 
has evolved into a versatile framework for performing computations within a wide
range of strongly coupled 
systems arising across theoretical physics. However, a proof of the original 
$\AdS_5\times \Srm^5$ conjecture is still lacking, hampered by our current lack of 
understanding of strongly coupled gauge theory on the one hand and of string theory 
on RR backgrounds on the other. As an intermediate step it is thus desirable to 
look for simpler versions of the correspondence that exhibit some of its features but 
bypass many of the complexities of gauge and string theory. 

Recently the higher-spin theories of Vasiliev on anti de Sitter space \cite{Vasiliev:2003ev,Vasiliev:1999ba} have 
received a lot of attention. These highly non-linear theories consist of
a tower of interacting massless higher-spin fields and are somewhere between conventional field theories and string theory
in terms of complexity. They  are believed to
be related to the tensionless limit of superstring theory (see \cite{Chang:2012kt} for some recent developments).

In \cite{Sezgin:2002rt} it was conjectured that Vasiliev's minimal bosonic theory on $\AdS_4$  
with suitable boundary conditions on the bulk scalar field is dual to the free 
theory of $N$ massless scalars in its $O(N)$-singlet sector in the large $N$ limit.
This was extended by Klebanov and Polyakov \cite{Klebanov:2002ja} to
the critical $O(N)$ vector model. Recent calculations, such as three-point functions 
\cite{Giombi:2009wh}, have provided 
non-trivial evidence for this conjecture. See \cite{Giombi:2012ms} for a recent review. 

An even simpler duality was recently proposed by Gaberdiel and Gopakumar \cite{Gaberdiel:2010pz}.
Motivated by the conjecture of Klebanov and Polyakov, together with the 
observation \cite{Campoleoni:2010zq,Henneaux:2010xg,Gaberdiel:2010ar} that the asymptotic 
symmetries of the higher-spin generalization of gravity on 
$\AdS_3$ lead to $\Wcal$ algebras on the boundary, they proposed an exact duality between
$\mathcal W_N$ minimal models realized as the WZW coset
  \begin{equation}
    \frac{\widehat{\mathfrak{su}}(N)_k\times\widehat{\mathfrak{su}}(N)_1}{\widehat{\mathfrak{su}}(N)_{k+1}}
  \end{equation}
and the bosonic truncation of Vasiliev higher-spin theory on $\AdS_3$ in the 't Hooft limit
  \begin{equation}
    0\leq\lim_{N,k\rightarrow\infty} \frac{N}{N+k}\leq 1 \quad \text{fixed}.
  \end{equation}
This duality is simple both due to the usual power of conformal symmetry in two dimensions, which is only enhanced because of the
higher-spin symmetry, and because tensor fields with spin greater than one on $\AdS_3$ do not have any bulk degrees of freedom.

Since the original proposal, a large number of tests have been carried out, regarding
partition functions \cite{Gaberdiel:2011zw}, higher-spin black hole backgrounds
\cite{Gutperle:2011kf,Kraus:2011ds,Gaberdiel:2012yb,Kraus:2012uf}, and correlation functions
\cite{Chang:2011mz,Ammon:2011ua,Papadodimas:2011pf,Chang:2011vka}. These investigations have led to slight
refinements of the original conjecture and a better understanding of the matching of
states between the 
bulk and boundary theory \cite{Gaberdiel:2011aa,Castro:2011iw,Gaberdiel:2012ku,Banerjee:2012aj,Perlmutter:2012ds}.
Recent reviews of the conjecture and the above developments can be found in \cite{Gaberdiel:2012uj,Ammon:2012wc}.

 There have also been extensions of the conjecture beyond the original class of minimal
model CFTs \cite{Ahn:2011pv,Gaberdiel:2011nt}, and proposals with $\Ncal=2$ \cite{Creutzig:2011fe}
and more recently $\Ncal=1$ supersymmetry \cite{Creutzig:2012ar}. The $\Ncal=2$
proposal of \cite{Creutzig:2011fe}, which we will review below, has already been
subjected to several precise checks, such as the large-$N$ matching of partition
functions \cite{Candu:2012jq}, detailed analysis of the symmetries 
\cite{Henneaux:2012ny,Hanaki:2012yf,Ahn:2012fz}, and more recently an analysis
of the symmetries at the quantum level \cite{Candu:2012tr} which revealed an interesting duality structure
of equivalent theories at different values of the parameters. For related work based on $\mathfrak{sl}(N|N-1)$ 
supergravity, see \cite{Tan:2012xi,Datta:2012km}.

 In order to put the proposal of \cite{Creutzig:2011fe} on even firmer ground,
it is important to move beyond the symmetries and spectrum and compare 
correlation functions on both sides of the duality. This will be our goal in this
paper. In particular, we compute three-point functions holographically, using the 
higher-spin theory on $\AdS_3$, as well as directly from the boundary CFT with
$\mathcal{SW}_\infty[\lambda]$ symmetry. Our three-point functions will be of a restricted 
type, involving two bosonic matter fields and one bosonic higher-spin field. Despite the restriction to
bosonic fields, the richer structure of the $\Ncal=2$ theory allows us to compute several types
of correlation functions not present in the non-supersymmetric theory. 
As an illustration of our results, let us display the following three-point function containing 
massive scalars and a bosonic higher-spin current not present in the non-supersymmetric theory:
  \begin{gather*}
     \begin{split}
	\average[\big]{\mathcal O^{\mathcal B}_{\Delta_+}(z_1,\bar z_1)\bar{\mathcal O}^{\mathcal B}_{\Delta_+}(z_2,\bar z_2)W^{s-}(z_3)} &= (-1)^{s-1}\frac{\Gamma^2(s)}{\Gamma(2s-1)}\,\frac{\Gamma(s-2\lambda+1)}{\Gamma(2-2\lambda)}\,\frac{s-1+2\lambda}{2s-1}\left(\frac{z_{12}}{z_{13}z_{23}}\right)^s\\
		&\quad\times\average[\big]{\mathcal O^{\mathcal B}_{\Delta_+}(z_1,\bar z_1)\bar{\mathcal O}^{\mathcal B}_{\Delta_+}(z_2,\bar z_2)}.
      \end{split}
  \end{gather*}
For this class of correlation functions, we find precise matching between the bulk
and boundary calculations, thus lending further support to the $\Ncal=2$ version of
the minimal-model/higher-spin correspondence. 

 The plan of this paper is as follows. In the following section we will review
the proposal of \cite{Creutzig:2011fe}, motivate a small modification of Vasiliev theory and calculate the scalar masses in this formalism. This will at the same time fix our notation
and conventions. In section \ref{section:OPE} we will establish the precise AdS/CFT dictionary for the higher-spin fields by, using the bulk theory, deriving operator product expansions of conserved currents of the dual boundary CFT. As a side product, this gives a holographic proof of the emergence of $\mathcal N=2$ $\mathcal SW_\infty[\lambda]$ near the $\AdS_3$ boundary and is by itself a consistency check of the duality. In section \ref{Bulk}, which forms the main technical part of
the paper, we will perform the holographic computation of three-point functions
from the higher-spin $\AdS_3$ theory. The corresponding CFT calculation is
performed in section \ref{Boundary}, where (as already mentioned) precise agreement
is found. We conclude with a discussion of future directions and open problems.

Furthermore, we have included two appendices. In appendix \ref{appendix:VasilievTheory}, we will give a 
lightning review of the full non-linear $\mathcal N=2$ Prokushkin-Vasiliev theory and its linearization
 which is used in this paper. Finally, appendix \ref{appendix:StructureConstantsOfHigherSpinAlgebra} contains 
explicit formulas for the structure constants of $\mathcal{SB}[\mu]$ and $\shs[\lambda]$ algebras,
 together with certain useful relations and properties used in the paper.

\paragraph{Note Added:} 
 During the completion of this article we became aware of the parallel work \cite{Creutzigetal},
which also considers three-point functions in the $\Ncal=2$ duality. That work
computes three-point functions with fermionic primaries and higher-spin bosonic
currents, which are not considered here, and achieves a better understanding of the 
relation between the bulk and CFT states. Although their bulk approach is similar to ours,
our boundary approaches are very different. Furthermore, the holographic OPE's of 
section \ref{section:OPE} are not considered in \cite{Creutzigetal}. Where there is overlap, 
we find agreement with the results of \cite{Creutzigetal}.  
\section[The \TorPDF{\Ncal = 2}{N=2} Minimal model -- Higher-spin duality]{The $\Ncal=2$ Minimal model -- Higher-spin duality} \label{Review}
In \cite{Creutzig:2011fe}, it was conjectured that the $\Ncal=(2,2)$ $\mathbb CP^N$ Kazama-Suzuki model \cite{Kazama:1988qp}, which can be represented as an ordinary coset \cite{Naculich:1997ic}
  \begin{equation}\label{eq:KazamaSuzukiCoset}
    \frac{\widehat{\mathfrak{su}}(N+1)_k\times\widehat{\mathfrak{so}}(2N)_1}{\widehat{\mathfrak{su}}(N)_{k+1}\times\widehat{\mathfrak{u}}(1)_{N(N+1)(k+N+1)}},
  \end{equation}
is dual to the $\mathcal N=2$ supersymmetric Prokushkin-Vasiliev theory \cite{Prokushkin:1998bq,Prokushkin:1998vn} on $\AdS_3$ with the parameter identification\footnote{Note that the parameter $\tilde\lambda$ used in \cite{Creutzig:2011fe} is related to ours by $\tilde\lambda = 2\lambda$.} $\lambda = \frac N{2(N+k)}$ in the 't Hooft limit
  \begin{equation}
      0\leq\lim_{N,k\rightarrow\infty}\lambda\leq \frac 12\quad \text{fixed}.
  \end{equation}
The notation $\hat{\mathfrak g}_k$ stands for the untwisted affine Lie algebra associated to $\mathfrak g$, at level $k$.\footnote{The factor $\widehat{\mathfrak{so}}(2N)_1$ arises due to the adjoint fermions of the affine Lie superalgebra. This factor appears since the coset is written in terms of ordinary affine Lie algebras, instead of affine Lie superalgebras. The first factor is associated to $\mathbb CP^N = \frac{SU(N+1)}{SU(N)\times U(1)}$.} Vasiliev theory contains two massive 3d $\mathcal N=2$ hypermultiplets, $\big(\phi_\pm, \psi_\pm\big)$ and  $\big(\tilde\phi_\pm, \tilde\psi_\pm\big)$, with two complex scalars and two fermions in each, with the masses
  \begin{equation}
   (M^B_+)^2 =  4(\lambda^2-\lambda), \qquad (M^B_-)^2 = 4\lambda^2 - 1, \qquad (M^F_\pm)^2 = (2\lambda-\tfrac 12)^2,
  \end{equation}
and equal masses for the other multiplet. The two multiplets have slightly different couplings to the massless higher-spin fields. The massless sector can be formulated as a $\shs[\lambda]_{k_{CS}}\times\shs[\lambda]_{-k_{CS}}$ Chern-Simons theory\footnote{In \cite{Prokushkin:1998vn} this Lie superalgebra is called $hs(2,\nu)\oplus hs(2,\nu)$.} \cite{Prokushkin:1998vn}, where the super higher-spin algebra $\shs[\lambda]$ is an infinite-dimensional Lie algebra which can be thought of as an analytical continuation of $\mathfrak{sl}(\lambda-1,\lambda)$ to non-integer $\lambda$ (see appendix \ref{appendix:StructureConstantsOfHigherSpinAlgebra} for more details).

The asymptotic symmetries of a $\mathfrak g$ Chern-Simons theory together with Brown-Henneaux-type boundary fall-off conditions \cite{Campoleoni:2010zq, Campoleoni:2011hg}, translate into the classical Drinfeld-Sokolov reduction \cite{Bouwknegt:1992wg} of $\mathfrak g$. In the case of pure gravity, $\mathfrak g=\mathfrak{sl}(2,\mathbb R)$, this leads to the Virasoro algebra, while for $\mathfrak g=\shs[\lambda]$ this leads to the non-linear $\mathcal{SW}_{\infty}[\lambda]$ \cite{Hanaki:2012yf} algebra. On the other hand, in \cite{Ito:1991wb} the chiral algebra of the coset \eqref{eq:KazamaSuzukiCoset} was shown to be related to quantum Drinfeld-Sokolov reduction of $A(N,N-1)=\mathfrak{sl}(N+1,N)$ with the principal embedding of $\mathfrak{sl}(2,\mathbb R)$, which is the $\mathcal N=2$ super $\Wcal$-algebra $\mathcal{SW}_N$. Recently it was shown that in the 't Hooft limit, the chiral algebra has the limit $\lim_{N,k\rightarrow\infty}\mathcal{SW}_N = \mathcal{SW}_{\infty}[\lambda]$ \cite{Candu:2012tr}, which is crucial for the duality to hold and for the calculations in this paper.

The restriction of the range of the parameter $0\leq\lambda\leq\frac 12$ leads to scalar masses with $-1\leq (M^B)^2\leq 0$. It is well known  \cite{Klebanov:1999tb} that for this mass range one can choose two different boundary conditions, with the ``usual''
quantization being the one with the largest value of the conformal dimension. Recall the usual AdS/CFT dictionary 
between masses and conformal dimensions of dual CFT operators
  \begin{equation}
    (M^B)^2 = \Delta(\Delta-2), \qquad (M^F)^2 = (\Delta-1)^2,
  \end{equation}
for massive scalars and spin $1/2$ fermions, respectively. The dual conformal weights are given by \cite{Creutzig:2011fe}
  \begin{gather}\label{eq:ReviewSectionCFTfieldsWeigts}
    \begin{split}
      \big(\Delta^B_+,\,\Delta^F_\pm,\,\Delta^B_-\big) = \big(2-2\lambda,\,\tfrac 32-2\lambda,\,1-2\lambda\big),\qquad
      \big(\tilde{\Delta}^B_+,\,\tilde{\Delta}^F_\pm,\,\tilde{\Delta}^B_-\big) = \big(2\lambda,\, \tfrac 12+2\lambda,\, 1+2\lambda\big).
    \end{split}
  \end{gather}
The bosonic operators in the first multiplet correspond to the $\phi_+$ scalar with the usual quantization and the 
$\phi_-$ scalar with the alternative quantization, while the quantizations are opposite in the second multiplet. 

Let $(\rho,s;\nu,m)$ label the states of the coset \eqref{eq:KazamaSuzukiCoset} up to field identifications due to outer automorphisms of the different factors in the coset. Here $\rho$ and $\nu$ are highest weights of $\mathfrak{su}(N+1)$ and $\mathfrak{su}(N)$, respectively, while $m\in\mathbb Z_{N(N+1)(k+N+1)}$. In the NS sector we have $s=0,2$. In \cite{Creutzig:2011fe}, it was proposed that the following holomorphic coset primary fields with chiral conformal weights
  \begin{gather}\label{eq:CosetChiralFieldConformalWeights}
    \begin{aligned}
      h(f,0;0,N) &=\lambda, & h(0,2;f,-N-1) &= \frac 12-\lambda,\\
      h(f,2;0,N) &= \lambda+\frac 12,	& h(0,0;f,-N-1) &= 1-\lambda,
    \end{aligned}
  \end{gather}
where $f$ is the fundamental representation, can be used to construct the dual fields \eqref{eq:ReviewSectionCFTfieldsWeigts} by gluing holomorphic and anti-holomorphic states as follows
  \begin{gather}\label{eq:ReviewSectionCFTPrimariesFirstMultiplet}
    \begin{aligned}
      \mathcal O^{\mathcal B}_{\Delta_+}&=(0,0;f,-N\!-\!1)\otimes (0,0;f,-N\!-\!1),&\quad \mathcal O^{\mathcal F}_{\Delta_+}&=(0,2;f,-N\!-\!1)\otimes (0,0;f,-N\!-\!1), \\
      \mathcal O^{\mathcal B}_{\Delta_-}&=(0,2;f,-N\!-\!1)\otimes (0,2;f,-N\!-\!1),& \quad \mathcal O^{\mathcal F}_{\Delta_-}&=(0,0;f,-N\!-\!1) \otimes (0,2;f,-N\!-\!1),
    \end{aligned}
  \end{gather}
and for the other multiplet
  \begin{gather}\label{eq:ReviewSectionCFTPrimariesSecondMultiplet}
    \begin{aligned}
      \tilde{\mathcal O}^{\mathcal B}_{\Delta_+}&=(f,0;0,N)\otimes (f,0;0,N),&\quad \tilde{\mathcal O}^{\mathcal F}_{\Delta_+}(z,\zbar)&=(f,0;0,N)\otimes (f,2;0,N), \\
      \tilde{\mathcal O}^{\mathcal B}_{\Delta_-}&=(f,2;0,N)\otimes (f,2;0,N), &\quad \tilde{\mathcal O}^{\mathcal F}_{\Delta_-}(z,\zbar)&=(f,2;0,N) \otimes (f,0;0,N). 
    \end{aligned}
  \end{gather}

 In the 't Hooft limit, the correlation functions we will be considering only depend on the higher-spin algebra $\shs[\lambda]$.  
Thus, in section \ref{BoundarySection} we will generate the corresponding highest-weight representations using a free-field
CFT having $\shs[\lambda]$ as a subalgebra. Our highest-weight representations will then be constructed in terms of free fields
such that they match the above coset primary fields.

\subsection{Modified Vasiliev Theory}\label{eq:ModifiedVasilievFormalism}
In this paper we will only consider the linearized Vasiliev theory, in which the matter fields propagate on a fixed higher-spin background on $\AdS_3$. This means that we will not take into account effects such as backreaction of matter fields on the higher-spin fields and non-linear interactions between the matter fields. See appendix \ref{appendix:VasilievTheory} for a very brief review of how this linearized theory comes out of the full non-linear Vasiliev theory.

The linearized Vasiliev theory is formulated in terms of two spacetime one-forms which can be identified with Chern-Simons gauge fields and describe the tower of higher-spin fields $A(\tilde y;k|x)$ and $\bar A(\tilde y;k|x)$, and two 0-forms which are generating functions of the matter fields, $C(\tilde y;k|x)$ and $\tilde C(\tilde y;k|x)$. They take values in the associative algebra $Aq(2,\nu)$ \cite{Vasiliev:1989qh} generated by the $\mathfrak{sl}(2,\mathbb R)$ spinor $\tilde y_\alpha$ and $k$ modulo the relations
  \begin{equation}\label{eq:GeneratingElementsOfAq(2,nu)}
    [\tilde y_\alpha,\tilde y_\beta]_\star = 2i\epsilon_{\alpha\beta}(1+\nu k), \qquad \{\tilde y_\alpha,k\} = 0, \qquad k^2=1.
  \end{equation}
The product between the so-called deformed oscillators $\tilde y_\alpha$ will be denoted by $\star$, and accordingly $[\cdot,\cdot]_\star$ and $\{\cdot,\cdot\}_\star$ will denote the commutator and anti-commutator wrt. to the $\star$-product, respectively. In this algebra the fields have the following expansions
  \begin{gather}\label{eq:ExpansionInAq(2,nu)}
    \begin{aligned}
      C(\tilde y;k|x) &= \sum_{B=0}^1\sum_{n=0}^\infty \frac 1{n!} C^B_{\alpha_1\dots\alpha_n}(x)\,k^B\,\tilde y^{\alpha_1}\star\dots\star\tilde y^{\alpha_n},\\
      A(\tilde y;k|x) &= \sum_{B=0}^1\sum_{n=0}^\infty \frac 1{n!} A^B_{\alpha_1\dots\alpha_n}(x)\,k^B\,\tilde y^{\alpha_1}\star\dots\star\tilde y^{\alpha_n},
    \end{aligned}
  \end{gather}
and similarly for $\bar A$ and $\tilde C$. The coefficients are symmetrized in the $\alpha$ indices and they have Grassmann parity equal to the number of indices mod 2. Thus commutators of elements in the algebra automatically turn into supercommutators of $\tilde y_\alpha$ and $k$ polynomials. In fact, as we will see in a moment, supercommutators of symmetrized elements of $Aq(2,\nu)$, with the above $\mathbb Z_2$-grading, form the infinite dimensional Lie superalgebra $hs(2,\nu) \approx \shs[\lambda]$ \cite{Prokushkin:1998vn}, with the identification $\lambda = \frac{\nu+1}4$.

The generating element $k$ is responsible for doubling the number of fields and thereby the $\mathcal N=2$ extension of the supersymmetry. The invariant subsets are projected out as
  \begin{equation}
    C = \Pi_+ C_+ +\Pi_- C_- \quad \text{and}\quad \tilde C =  \Pi_+ \tilde{C}_++\Pi_-\tilde{C}_-\;,
\quad \text{where}\quad \Pi_\pm = \frac{1\pm k}2.
  \end{equation}
The lowest components $\phi_\pm\equiv C^\pm_0$ and $\psi_\pm \equiv C^\pm_\alpha$ correspond to the two complex scalars and two fermions discussed in the previous section, respectively, while the fields with more that two spinor indices form a tower of auxiliary fields. There are four corresponding fields from $\tilde C^\pm$ and all together we have two sets of $3d$ $\mathcal N=2$ hypermultiplets
  \begin{equation}
    (\phi_+,\psi_+,\psi_-,\phi_-),\qquad \text{and}\qquad (\tilde \phi_+,\tilde \psi_+,\tilde \psi_-,\tilde \phi_-). 
  \end{equation}
The linearized Vasiliev equations for the matter fields are (see appendix \ref{appendix:VasilievTheory})
  \begin{gather}\label{eq:VasilievEquationMatterCoupledToGaugeField}
    \begin{aligned}
      \ud C+A\star C - C\star\bar A &=0,\\
      \ud\tilde C + \bar A\star\tilde C-\tilde C\star A &= 0,
    \end{aligned}
  \end{gather}
while the equations for the one-forms are just flatness conditions
    \begin{equation}\label{eq:VasilievEquationsOfMotionForGaugeFields}
    \ud A + A\star\wedge A = 0,\qquad \ud\bar A + \bar A\star\wedge\bar A = 0.
  \end{equation}
Note that the flatness conditions only involve (anti-)commutators when written in component form, so if we turn off the matter fields the theory reduces to a $\shs[\lambda]_{k_{CS}}\times\shs[\lambda]_{-k_{CS}}$ Chern-Simons theory which is the $\mathcal N=2$ higher-spin SUGRA recently studied in \cite{Hanaki:2012yf}. The full associative algebra $Aq(2,\nu)$ only enters through coupling to matter fields as seen from the equations \eqref{eq:VasilievEquationMatterCoupledToGaugeField}.

It is obvious that this formalism quickly becomes very tedious. We have to multiply pairs of symmetrized elements of $Aq(2,\nu)$, then use the relations \eqref{eq:GeneratingElementsOfAq(2,nu)} to express the result as sums of symmetrized products of $\tilde y_\alpha$ and $k$. Inspired by the calculation in \cite{Ammon:2011ua}, we would like to have closed-form expressions for the products. This is what we will consider now.

The Lie algebra structure $\shs[\lambda]$ can be inherited from the associative product of an algebra we will call $\mathcal{SB}[\mu]$, which can be constructed as the following quotient \cite{Bergshoeff:1991dz}
  \begin{equation}\label{eq:SBmuAsAQuotientOfUniversalEnvelopingAlgebraOfOSP12}
    \mathcal{SB}[\mu] = \frac{U(\mathfrak{osp}(1|2))}{\langle\mathcal C_2-\mu\mathbbm 1\rangle} = \shs[\lambda]\oplus\mathbb C,
  \end{equation}
where $U(\mathfrak{osp}(1|2))$ is the universal enveloping algebra of $\mathfrak{osp}(1|2)$, $\mathcal C_2$ is its second order Casimir element, $\mu = \lambda(\lambda -\frac 12)$ and the factor $\mathbb C$ is spanned by the identity element of $\mathcal{SB}[\mu]$ (see more details in appendix \ref{appendix:StructureConstantsOfHigherSpinAlgebra}).\footnote{See \cite{Gaberdiel:2011wb} for a discussion of the non-supersymmetric construction.} The Lie algebra $\shs[\lambda]$ contains the following infinite tower of generators, 
  \begin{gather}\label{eq:ListOfHigerSpinAlgebraGenerators}
      \begin{aligned}
	  &L^{(s)+}_m \quad(s\in\mathbb Z_{\geq 2}, |m|\leq s-1),\quad &\quad&L^{(s)-}_m \quad(s\in\mathbb Z_{\geq 1}, |m|\leq s-1),\\
	  &G^{(s)+}_r \quad(s\in\mathbb Z_{\geq 2}, |r|\leq s-3/2),\quad &\quad&G^{(s)-}_r\quad(s\in\mathbb Z_{\geq 2}, |r|\leq s-3/2),
      \end{aligned}
  \end{gather}
where $m\in\mathbb Z$ while $r\in\mathbb Z+\frac 12$. In addition to these, $\mathcal{SB}[\mu]$ contains the identity element which we will write as $L^{(1)+}_m\equiv\mathbbm 1$. The generators $L_m \equiv L^{(2)+}_m$ and $G_\alpha\equiv G^{(2)+}_\alpha$ form an $\mathfrak{osp}(1|2)$ subalgebra, where $G_\alpha$ is the supercharge. Actually we also get a $\mathfrak{osp}(2|2)$ subalgebra if we add the other supercharge $G^{(2)-}_r$ and the generator of $U(1)$ R-symmetry $L^{(1)-}_0$. According to \eqref{eq:SBmuAsAQuotientOfUniversalEnvelopingAlgebraOfOSP12} we can express all the generators \eqref{eq:ListOfHigerSpinAlgebraGenerators} in terms of the $\mathfrak{osp}(1|2)$ generators, but $G_\alpha$ is actually sufficient since all $L_m$ can be written as anti-commutators of $G_\alpha$.

It is shown in \cite{Bergshoeff:1991dz} that $\mathcal{SB}[\mu]$ can be generated by $G_\alpha$ together with an element $K$ (which is essentially the commutator of $G_\alpha$) with the properties
  \begin{equation}\label{eq:GeneratingElementsOfSB[mu]}
    [G_\alpha, G_\beta] = (cK+\tfrac 12)\epsilon_{\alpha\beta},\qquad \{K,G_\alpha\} = 0, \qquad K^2 = 1,
  \end{equation}
where $c = 2(\lambda -\frac 14)$. Using the identifications $\nu = 2c = 4\lambda-1$ and
  \begin{equation}
    G_\alpha = \left(\frac {-i}4\right)^{1/2}\tilde y_\alpha\;,
  \end{equation}
we clearly see that \eqref{eq:GeneratingElementsOfSB[mu]} and \eqref{eq:GeneratingElementsOfAq(2,nu)} are equivalent. Thus $Aq(2,\nu)$ is isomorphic to $\mathcal{SB}[\mu]$. We can actually directly write down the all the generators \eqref{eq:ListOfHigerSpinAlgebraGenerators} in terms of $\tilde y_\alpha$ and $k$. By looking at the (anti-)commutators of the $\mathfrak{osp}(1|2)$ and $\mathfrak{osp}(2|2)$ together with appendix B of \cite{Ammon:2011ua}, it is clear that the $\shs[\lambda]$ generators are related to the $Aq(2,\nu)$ generators by
\be
 L_m^{(s)+}\sim \left(\frac {-i}4\right)^{s-1}S^s_m\; \qquad\text{and}\qquad 
L_m^{(s)-}\sim \left(\frac {-i}4\right)^{s-1}S^s_m k\;,
\ee
for integer $s$ and 
\be
G_m^{(s+\half)+}\sim \left(\frac {-i}4\right)^{s-1}S^s_m\; \qquad\text{and}\qquad 
 G_m^{(s+\half)-}\sim \left(\frac {-i}4\right)^{s-1}S^s_m k\;
\ee
for half-integer $s$.
Here $S^s_m$ is a symmetric product of $2(s-1)$ $\tilde y_\alpha$'s with $N_\pm$ of $y_{\pm \frac 12}$ and $2m = N_+-N_-$. We will however not need the explicit mapping between $\mathcal{SB}[\mu]$ and $Aq(2,\nu)$ in this paper, only the fact that they are isomorphic.

In \cite{Bergshoeff:1991dz} the structure constants of the (linear) $\Wcal$-algebra $sw_\infty[\lambda]$ and an associative extension thereof were explicitly constructed. It turns out that $\mathcal{SB}[\mu]$ is a subalgebra of this. By some work we can extract the structure constants in the form which is convenient for us\footnote{This was also recently discussed in the appendix of \cite{Hanaki:2012yf}.}, see appendix \ref{appendix:StructureConstantsOfHigherSpinAlgebra} for notation and concrete formulas.

We will hereby modify the traditional Vasiliev formalism by changing $Aq(2,\nu)$ into $\mathcal{SB}[\mu]$. It is convenient to simplify the notation by allowing $s$ to be half-integer and identifying
  \begin{equation}
      L^{(s)}_m \equiv L^{(s)+}_m,\quad L^{(s+1/2)}_m \equiv L^{(s)-}_m,\quad G^{(s)}_r \equiv G^{(s)+}_r\quad\text{and}\quad G^{(s-1/2)}_r \equiv G^{(s)-}_r.
  \end{equation}
In this formalism the expansions \eqref{eq:ExpansionInAq(2,nu)} of the generating functions are given as
  \begin{gather}
    \begin{split}\label{eq:ExpansionOfGaugeAndMatterFieldInTermsOfSuperHigherSpinAlgebra}
      A &= \sumCircles_{s=1}^{\infty}\sum_{|m|\leq s-1}A_m^s\:L_m^{(s)}+ \sumCircles_{s = \frac 32}^{\infty}\sum_{|r|\leq s-\frac 32} A_r^s\:G_r^{(s)},\\
      C &= \sumCircles_{s=1}^{\infty}\sum_{|m|\leq s-1}C_m^s\:L_m^{(s)} + \sumCircles_{s = \frac 32}^{\infty}\sum_{|r|\leq s-\frac 32} C_r^s\:G_r^{(s)},
    \end{split}
  \end{gather}
and similarly for $\tilde C$ and $\bar A$. The notation $\sumCircles$ stands for summation over half-integer steps. Note that we can easily distinguish the bosonic components $C^s_m$ from the fermionic (anticommuting) ones $C^s_r$, since $m$ is always an integer while $r$ is half of an odd integer. In this formalism the physical scalars $\phi_\pm$ and fermions $\psi_\pm$, are given by appropriate superpositions of the lowest components $C^1_0$, $C^{\frac 32}_0$, $\Big\{C^{\frac 32}_{+\frac 12},C^{\frac 32}_{-\frac 12}\Big\}$ and $\Big\{C^2_{+\frac 12},C^2_{-\frac 12}\Big\}$.

In appendix \ref{appendix:VasilievTheory} we mention the fact that, using the projection operator $\Pi_\pm = \frac{1\pm k}2$,
 the bosonic subalgebra of $Aq(2,\nu)$ decomposes into $Aq^E(2,\nu)\oplus Aq^E(2,-\nu)$, which is isomorphic to 
$\hs[\frac{1-\nu}2]\oplus\hs[\frac{1+\nu}2]$ . Since the same projector is also used to extract $\phi_\pm,\psi_\pm$, when 
computing the bosonic three-point function we only need the subalgebra given by $\hs[\frac{1\mp \nu}2]$. Therefore, the
three-point functions could be extracted from the results of \cite{Ammon:2011ua} using the relation $\nu=4\lambda-1=1-2\lambda_{AKP}$,
where $\lambda_{AKP}$ is the parameter appearing in \cite{Ammon:2011ua}. However, in the fermionic case, we need to use more than
just the bosonic subalgebra, and we only know the coefficients in the basis \eqref{eq:ListOfHigerSpinAlgebraGenerators}. In
this basis the fermions and bosons do not come out as naturally, so with a view to extending our results to eventually include
fermions we choose to perform the full bosonic calculation in this basis.

\subsection[Scalars Propagating on \TorPDF{\AdS_3}{Ads3}]{Scalars Propagating on $\AdS_3$}\label{Subsection:ScalarsOnAdS3}
In this section we will illustrate how the Vasiliev equations \eqref{eq:VasilievEquationMatterCoupledToGaugeField} give rise to the Klein-Gordon equation on $\AdS_3$ for the scalars, with the correct masses as known in the literature \cite{Prokushkin:1998bq, Creutzig:2011fe}. In the traditional formalism of Vasiliev based on the deformed oscillators $\tilde y_\alpha$ and $k$, the same calculation would be much more tedious.

The connection corresponding to $\AdS_3$ is given as
  \begin{gather}\label{eq:AdS3GaugeFieldsAndMetric}
    \begin{split}
	A &= e^{\rho}\, L^{(2)}_1\,\ud z +  L^{(2)}_0\,\ud\rho\\
	\bar A &= e^{\rho}\, L^{(2)}_{-1}\,\ud\bar z - L^{(2)}_0\,\ud\rho
    \end{split}
     \qquad\Rightarrow\qquad
    \ud s^2 = \ud\rho^2 + e^{2\rho}\,\ud z\ud\bar z,
  \end{gather}
where we have mapped to the metric formulation by $g_{\mu\nu} = \frac 12\text{tr}(e_{\mu}e_{\nu})$, $e = \frac 12\left(A - \bar A\right)$ \cite{Campoleoni:2010zq,Campoleoni:2011hg}. The trace is defined and normalized as follows
  \begin{gather}
    \begin{split}
      \text{tr}\big(\mathcal A\star\mathcal B\big) &= \frac{\mathcal A\star\mathcal B}{\left(2\lambda^2-\lambda\right)}\bigg|_{\mathcal J = 0},\qquad\forall\mathcal J\neq\mathbbm 1.
    \end{split}
  \end{gather}
Turning on other modes, such that \eqref{eq:VasilievEquationsOfMotionForGaugeFields} and appropriate boundary conditions are satisfied, corresponds to higher-spin deformations of $\AdS_3$. We will for now only consider the scalar fields propagating on $\AdS_3$, so we will set the fermionic coefficients $C^s_r = 0$. Plugging \eqref{eq:ExpansionOfGaugeAndMatterFieldInTermsOfSuperHigherSpinAlgebra} into Vasiliev equation \eqref{eq:VasilievEquationMatterCoupledToGaugeField} we find
  \begin{gather}
    \begin{split}
      \sumCircles_{s=1}^{\infty}\sum_{|m|\leq s-1}\bigg(\ud C^s_m\, L^{(s)}_m + e^{\rho}\,C^s_m\,L^{(2)}_1&\star L^{(s)}_m\,\ud z - e^{\rho}\,C^s_m\, L^{(s)}_m\star L^{(2)}_{-1}\,\ud\bar z\\
	&+\,C^s_m\,\Big\{L^{(2)}_0\star L^{(s)}_m + L^{(s)}_m\star L^{(2)}_0\Big\}\,\ud\rho \bigg) = 0.
    \end{split}
  \end{gather}
The coefficients of linearly independent terms should be set to zero individually. Using the properties of the structure constants given in appendix \ref{appendix:StructureConstantsOfHigherSpinAlgebra}, we find the following set of coupled equations
  \begin{gather}\label{eq:VasilievEquationRecursionRelationsForScalars}
    \begin{aligned}
      &\partial_{\rho}C^s_m + 2\,\bigg[\,C^{s-1}_m +\,C^{s+1}_m\,g^{s+1,2}_3(m,0) +\,C^{s-\frac 12}_m\,g^{s-\frac 12,2}_{\frac 32}(m,0) +\,C^{s+\frac 12}_m\,g^{s+\frac 12,2}_{\frac 52}(m,0)\bigg] = 0,\\
      &\begin{aligned}
	  \partial C^s_m + e^{\rho}\,\bigg[C^{s-1}_{m-1} + g^{2,s}_2(1,m-1)&C^s_{m-1} + g^{2,s+1}_3(1,m-1)C^{s+1}_{m-1}\\
			      + & g^{2,s-\frac 12}_{\frac 32}(1,m-1)C^{s-\frac 12}_{m-1} + g^{2,s+\frac 12}_{\frac 52}(1,m-1)C^{s+\frac 12}_{m-1}\bigg] = 0,
	\end{aligned}\\
      &\begin{aligned}
	\bar{\partial}C^s_m - e^{\rho}\,\bigg[C^{s-1}_{m+1} + g^{s,2}_2(m+1&,-1)C^s_{m+1}+g^{s+1,2}_3(m+1,-1)C^{s+1}_{m+1}\\
				+& g^{s-\frac 12,2}_{\frac 32}(m+1,-1)C^{s-\frac 12}_{m+1} + g^{s+\frac 12,2}_{\frac 53}(m+1,-1)C^{s+\frac 12}_{m+1}\bigg] = 0.
	\end{aligned}
    \end{aligned}
  \end{gather}
Note that we obviously define $C^s_m = 0$ for modes outside of the wedge $|m|>s-1$. These equations can be solved recursively in order to express the auxiliary fields in terms of $C^1_0$ and $C^{\frac 32}_0$, and thereby find the equations of motion of these scalars. Analyzing the structure of these equations, it turns out that the minimal number of equations needed are
  \begin{alignat*}{2}
    &L_{0,\rho}^{(1)}:\hspace{6mm}	&	&\partial_{\rho}C_0^1 + \lambda(2\lambda -1)C_0^2 = 0,\\
    &L_{0,\rho}^{(\frac 32)}:&	&\partial_{\rho}C_0^{\frac 32}+\frac 19(2\lambda^2-\lambda-1)C_0^{\frac 52}+\frac 16(4\lambda-1)C_0^2=0,\\
    &L_{0,\rho}^{(2)}:	&	&\partial_{\rho}C_0^2 + 2C_0^1 + \frac 23(1-4\lambda)C_0^{\frac 32}+\frac 49(2\lambda^2-\lambda-1)C_0^3=0,\\
    &L_{0,\rho}^{(\frac 52)}:&	&\partial_{\rho}C_0^{\frac 52}+2C_0^{\frac 32} + \frac 2{15}(4\lambda-1)C_0^3+\frac 4{25}(2\lambda^2-\lambda-3)C_0^{\frac 72}=0,\\
    &L_{0,\bar z}^{(1)}:	&	&\bpartial C_0^1 - e^{\rho}(1-2\lambda)\lambda C_1^2=0,\\
    &L_{0,\bar z}^{(\frac 32)}:&	&\bpartial C_0^{\frac 32} - e^{\rho}\left[\frac 16(1-4\lambda)C_1^2-\frac 19(1+\lambda-2\lambda^2)C_0^{\frac 52}\right]=0,\\
    &L_{1, z}^{(2)}: &		&\partial C_1^2 + e^{\rho}\left[C_0^1+\frac 12 C_0^2+ \frac 19(1+\lambda-2\lambda^2)C_0^3 + \frac 13(1-4\lambda)C_0^{\frac 32}\right]=0,\\
    &L_{1,z}^{(\frac 52)}:	&	&\partial C_1^{\frac 52}+e^{\rho}\left[C_0^{\frac 32} + \frac 12C_0^{\frac 52}+\frac 1{25}(3+\lambda-2\lambda^2)C_0^{\frac 72}+\frac 1{30}(1-4\lambda)C_0^3\right]=0.
  \end{alignat*}
Solving these recursively we can eliminate all the auxiliary fields and reduce to two coupled equations
  \begin{gather}
    \begin{aligned}
      &\Box C_0^1 + 6\lambda\, (1-2 \lambda)\,C_0^1 + 2\lambda\,(1-6\lambda+8\lambda^2)\,C_0^{3/2} &= 0,&\\
      &\Box C_0^{3/2} - \frac{1-4\lambda}{6\lambda(1-2\lambda)}\,\Box C_0^1  + \frac 23(1+\lambda-2\lambda^2)\,C_0^{3/2} &=0,&
    \end{aligned}
  \end{gather}
with the Laplacian of $\AdS_3$ in the coordinates \eqref{eq:AdS3GaugeFieldsAndMetric} given by
  \begin{equation}
    \Box = \partial_{\rho}^2 + 2\,\partial_{\rho} + 4\, e^{-2\rho}\,\partial\bar{\partial}.
  \end{equation}
In order to bring these equations in standard form, we can remove the $\Box C^1_0$ term of the second equation by subtracting these two equations with an appropriate weight. This leads to the coupled Klein-Gordon equations
\begin{equation}
  \Box\bm C + \begin{bmatrix}
	      6\lambda(1-2\lambda)	&	2\lambda(1-6\lambda+8\lambda^2)\\
		1-4\lambda		&	1-2\lambda+4\lambda^2
           \end{bmatrix}
      \bm C = 0, \qquad 
        \bm C = \begin{pmatrix}
	  C_0^1 \\ C_0^{\frac 32}
        \end{pmatrix}.
\end{equation}
The fields $C^1_0$ and $C^{\frac 32}_0$ are clearly not ``mass-eigenstates'', but their superpositions must be. Diagonalizing the mass matrix we find
  \begin{gather}
    \begin{aligned}
      \Big[\Box - 4\,\left(\lambda^2-\lambda\right)\Big]\,\phi_+ = 0,\qquad
      \Big[\Box - \left(4\,\lambda^2-1\right)\Big]\,\phi_- = 0.
    \end{aligned}
  \end{gather}
Thus the masses of the two scalars are given by 
  \begin{equation}
      \qquad (M_+^B)^2=  4(\lambda^2-\lambda)\qquad\text{and}\qquad(M_-^B)^2 = 4\lambda^2-1,
  \end{equation}
and from the eigenvectors of the mass matrix we read off the correct superpositions
  \begin{equation}
    C^1_0 = (2 \lambda -1)\:\phi_+ + 2\:\lambda\:\phi_-, \hspace{10mm} C^{\frac 32}_0 = \phi_+ + \phi_-. \label{eq:VasilievLowestComponentScalarAsSuperpositionsOfPhysicalScalars}
  \end{equation}
By rescaling $\lambda = \frac 12\tilde{\lambda}$, the masses $(M_-^B)^2 = \tilde{\lambda}^2-1$ and $(M_+^B)^2 = \tilde{\lambda}^2-2\tilde{\lambda}$, exactly match the results known from the traditional Vasiliev theory \cite{Creutzig:2011fe,Prokushkin:1998bq}.
This confirms that our formulation works as expected without the very tedious manipulations involved in the  deformed oscillator approach. The advantages of having explicit formulas for the structure constants of $\mathcal{SB}[\mu]$ cannot be understated: without them our approach, originally laid out in \cite{Ammon:2011ua}, would be very hard to use for extracting three-point functions for arbitrary spin $s$.

With higher-spin deformations of $\AdS_3$, one can show that the Klein-Gordon equations get higher derivative corrections, as also observed in \cite{Ammon:2011ua}. We will however not need any of these in this paper.
\section{Holographic OPE's and the AdS/CFT dictionary}\label{section:OPE}

Recall that the Brown-Henneaux type asymptotic fall-off conditions \cite{Campoleoni:2010zq,Campoleoni:2011hg} translate into classical Drinfeld-Sokolov reduction of the gauge algebra $\shs[\lambda]_k\times\shs[\lambda]_{-k}$ with respect to the $\mathfrak{sl}(2,\mathbb R)$ embeddings, which correspond to a set of first class constraints. In the so-called lowest-weight gauge \cite{Bouwknegt:1992wg,Campoleoni:2011hg,Hanaki:2012yf}, the super-connections of constant $\rho$ slices take the form
  \begin{gather}
    \begin{aligned}
      a(z) &= \Bigg(L^{(2)}_1 + \frac{2\pi}k\,\sumCircles_{s\geq\frac 32}\left[\frac 1{N^B_s}\,\mathcal L_s\, L^{(s)}_{-\floor s+1} + \frac 1{N^F_s}\,\psi_s\, G^{(s)}_{-\ceil s+\frac 32}\right]\Bigg)\,\ud z,\\
      \bar a(\bar z) &= \Bigg(L^{(2)}_{-1} + \frac{2\pi}k\,\sumCircles_{s\geq\frac 32}\left[\frac 1{N^B_s}\,\bar{\mathcal L}_s\, L^{(s)}_{\floor s-1} + \frac 1{N^F_s}\,\bar{\psi}_s\, G^{(s)}_{\ceil s-\frac 32}\right]\Bigg)\,\ud\bar z,
    \end{aligned}
  \end{gather}
where only terms of lowest mode are allowed. $\floor s$ and $\ceil s$ are the floor and ceiling operators. Here $k = \frac l{4G}$ is the Chern-Simons level\footnote{Not to be confused with the $Aq(2,\nu)$ generator $k$ in \eqref{eq:GeneratingElementsOfAq(2,nu)}.} and the factor $2\pi/k$ is extracted such that the above solution reduces to the BTZ black hole when turning off higher-spin contributions. We choose the normalizations as
  \begin{equation}\label{eq:NormalizationOfSourceTermsUsingTrace}
    N^B_s = -\text{tr}\left(L^{(s)}_{-\floor s+1}L^{(s)}_{\floor s-1}\right),\qquad N^F_s = \text{tr}\left(G^{(s)}_{\ceil s-\frac 32}G^{(s)}_{-\ceil s+\frac 32}\right).
  \end{equation}
The functions $\mathcal L_s$ and $\psi_s$ must be holomorphic while $\bar{\mathcal L}_s$ and $\bar{\psi}_s$ must be anti-holomorphic in order to solve the equations of motion \eqref{eq:VasilievEquationsOfMotionForGaugeFields}. In order to calculate correlation functions \cite{Witten:1998qj,Gubser:1998bc} containing a holomorphic field of spin $s$, we need to add a corresponding source term to the boundary CFT action
  \begin{equation}\label{eq:BoundaryActionAddedSourceTerm}
    S_\partial \rightarrow S_\partial -\int\ud^2 z\,\mu_s(z)\,W^s(z).
  \end{equation}
Note that the spin $s$ field $W^s$ is irrelevant in the renormalization group sense and will therefore change the UV-structure of the dual CFT, which from the bulk perspective corresponds to that the geometry will no longer asymptote to the same $\AdS_3$ geometry.

From the standard prescription of AdS/CFT, the source terms correspond to boundary values of the dual bulk-fields, therefore we need to generalize the boundary conditions. Inspired by the spin-3 case \cite{Gutperle:2011kf}, we propose the following generalization of the super-connection
  \begin{gather}\label{eq:GeneralizedAnsatzForGaugeFieldSinceWeNeedBoundarySourceTerms}
    \begin{aligned}
      a = \Bigg(L^{(2)}_1 + \frac{2\pi}k\,&\sumCircles_{s\geq\frac 32}\left[\frac 1{N^B_s}\,\mathcal L_s\, L^{(s)}_{-\floor s+1} + \frac 1{N^F_s}\,\psi_s\, G^{(s)}_{-\ceil s+\frac 32}\right]\Bigg)\,\ud z\\
	& +\Bigg(\sumCircles_{s\leq\frac 32}\sum_{|m|\leq\floor s-1}\mu^s_m\, L^{(s)}_m + \sumCircles_{s\leq\frac 32}\sum_{|r|\leq\ceil s -\frac 32}\nu^s_r\, G^{(s)}_r\Bigg)\,\ud\bar z,
    \end{aligned}
  \end{gather}
where the functions $\mu^s_m = \mu^s_m(z,\bar z)$ and $\nu^s_r = \nu^s_r(z,\bar z)$ are non-chiral functions.\footnote{In most of the paper we will concentrate on the holomorphic sector, the discussion of the other sector is analogous.} Following the ideas developed in \cite{Gutperle:2011kf}, we will show that evaluating the bulk equations of motion to this ansatz, will yield the Ward identities of the dual CFT in the presence of higher-spin sources. We can in particular show the emergence of $\mathcal N=2\;\mathcal SW_{\infty}[\lambda]$ symmetry near the $\AdS_3$ boundary, by deriving the OPE's of the conserved currents in the dual CFT using the bulk theory. This can be thought of as an alternative (and holographic) approach to probing the asymptotic symmetries, which was done in \cite{Hanaki:2012yf} using different means.

The full gauge field is given by
  \begin{gather}\label{eq:FullGaugeFieldWithRhoDependence}
    \begin{split}
      A &= b^{-1} a b + b^{-1}\ud b,\\
      \bar A &= b\bar ab^{-1} + b\ud b^{-1},
    \end{split}
    \qquad \text{where}\quad 
      b = e^{\rho\, L^{(2)}_0}.
  \end{gather}
Using the Baker-Campbell-Hausdorff formula
\[ e^{X}Y e^{-X} = e^{\text{ad}X} Y =Y+\left[X,Y\right]+\frac{1}{2!}[X,[X,Y]]+\frac{1}{3!}[X,[X,[X,Y]]]+\cdots, \]
and the fact that $L^{(2)}_0$ is $\text{ad}$-diagonalized in the basis \eqref{eq:ListOfHigerSpinAlgebraGenerators}
\[[L_0^{(2)},L_m^{(s)}] = -m\,L_m^{(s)}, \qquad [L_0^{(2)},G^{(s)}_r] = -r\, G^{(s)}_r,\]
we find the following $\rho$ dependence for each generator in \eqref{eq:FullGaugeFieldWithRhoDependence}
    \begin{equation}
	e^{-\rho L_0^{(2)}}\:L_m^{(s)}\:e^{\rho L_0^{(2)}} = L_m^{(s)}\:e^{m\rho}, \qquad e^{-\rho L_0^{(2)}}\:G_r^{(s)}\:e^{\rho L_0^{(2)}} = G_r^{(s)}\:e^{r\rho}.
    \end{equation}
This implies that terms with highest possible modes, $\mu^s_{\floor s-1}$ and $\nu^s_{\ceil s -\frac 32}$, are the most dominant near the boundary and can thus be regarded as source terms. Note that this is nothing but a Fefferman-Graham expansion of $A$, which happens to be finite. Thus in order to establish the AdS/CFT dictionary, we need to investigate if these terms can be identified with the sources in the boundary action \eqref{eq:BoundaryActionAddedSourceTerm}. It turns out they actually can be identified with the boundary sources \eqref{eq:BoundaryActionAddedSourceTerm}, up to a factor of $2\pi$.

\subsection{Flatness Conditions}
Using the ansatz \eqref{eq:GeneralizedAnsatzForGaugeFieldSinceWeNeedBoundarySourceTerms} and the equations of motion, we can collect all the terms into coefficients of the Lie algebra generators
  \begin{gather}
    \begin{split}
      \partial a_{\bar z}-\bar{\partial}a_z +[a_z,a_{\bar z}] = \sumCircles_{s\geq\frac 32}\bigg[\sum_{|m|\leq\floor s-1}c^B_{s,m}\,L^s_m + \sum_{|r|\leq \ceil s-\frac 32}c^F_{s,r}\,G^s_r\bigg] = 0,
    \end{split}
  \end{gather}
giving rise to the following two set of equations
  \begin{equation}\label{eq:EquationsToBeSolvedRecursivelyForTheSourceTermsAndThenToFindTheWardIdentities}
	c^B_{s,m} = 0 \qquad \text{and} \qquad c^F_{s,r} = 0.
  \end{equation}
The coefficients for the bosonic generators are found to be
   \begin{gather}
      \begin{aligned}
	c^B_{s,m} &= \partial\mu^s_m - \frac{2\pi}k\frac 1{N^B_s}\,\bar{\partial}\mathcal L_s\,\delta_{m,\negative\floor s+1} + \big(\floor s-m\big)\mu^s_{m-1} \left(1-\delta_{m,\negative\floor s+1}\right)\\
	      &\quad+\frac{2\pi}k\sumCircles_{t\geq\frac 32}\Bigg\{\frac 1{N^B_t}\,\mathcal L_t\sumCircles_{\tilde s\geq\frac 32}\chi_{\big[-\floor{\tilde s}-\floor t+2,\floor{\tilde s}-\floor t\big]}(m)\,\mu^{\tilde s}_{m+\floor t-1}\sumCircles_{u=1}^{\tilde s+t-|\tilde s-t|-1}\delta_{\tilde s+t-u,s}                                                                                                                                                                                                                                                                                                                                                                                                                                                                                                                                                                                                                                                                                                                                                                                                                                                                                                                                                                                                                                                                                                                                                                                                                                                                                                                                                                                                                                                                                                                                                                                                                                                                                                                                                                                                                                                                                                                                                                                                                                                                                                                                                                                                                                                                                                                                                                                                                                                                                                                                                                         \\& \hspace{80mm}\times \hat g^{t,\tilde s}_u\Big(\negative\floor t+1,m+\floor t-1;\lambda\Big)
	  \\&\quad+ \frac 1{N^F_t}\,\psi_t \sumCircles_{\tilde s\geq\frac 32}\chi_{\big[-\ceil{\tilde s}-\ceil t+3,\ceil{\tilde s}-\ceil t\big]}(m)\,\nu^{\tilde s}_{m+\ceil t-\frac 32}\sumCircles_{u=1}^{\tilde s+t-|\tilde s-t|-1}\delta_{\tilde s+t-u,s}
	    \\&\hspace{75mm}\times\hat{\tilde g}^{t,\tilde s}_u\Big(\negative\ceil t+\frac 32,m+\ceil t-\frac 32;\lambda\Big)\Bigg\},
      \end{aligned}
   \end{gather}
and for the fermionic generators we have
   \begin{gather}
      \begin{aligned}
	c^F_{s,r} &= \partial\nu^s_r - \frac{2\pi}k\frac 1{N^F_s}\,\bar{\partial}\psi_s\,\delta_{r,\negative\ceil s+\frac 32} + \big(\ceil s-\frac 12-r\big)\nu^s_{r-1} \left(1-\delta_{r,\negative\ceil s+\frac 32}\right)\\
	      &\quad+\frac{2\pi}k\sumCircles_{t\geq\frac 32}\Bigg\{\frac 1{N^B_t}\,\mathcal L_t\sumCircles_{\tilde s\geq\frac 32}\chi_{\big[-\ceil{\tilde s}-\floor t+\frac 52,\ceil{\tilde s}-\floor t-\frac 12\big]}(r)\,\nu^{\tilde s}_{r+\floor t-1}\sumCircles_{u=1}^{\tilde s+t-|\tilde s-t|-1}\delta_{\tilde s+t-u,s}                                                                                                                                                                                                                                                                                                                                                                                                                                                                                                                                                                                                                                                                                                                                                                                                                                                                                                                                                                                                                                                                                                                                                                                                                                                                                                                                                                                                                                                                                                                                                                                                                                                                                                                                                                                                                                                                                                                                                                                                                                                                                                                                                                                                                                                                                                                                                                                                                                                                                                                                                                         \\& \hspace{80mm}\times \hat h^{t,\tilde s}_u\Big(\negative\floor t+1,r+\floor t-1;\lambda\Big)
	  \\&\quad+ \frac 1{N^F_t}\,\psi_t \sumCircles_{\tilde s\geq\frac 32}\chi_{\big[-\floor{\tilde s}-\ceil t+\frac 52,\floor{\tilde s}-\ceil t+\frac 12\big]}(r)\,\nu^{\tilde s}_{r+\ceil t-\frac 32}\sumCircles_{u=1}^{\tilde s+t-|\tilde s-t|-1}\delta_{\tilde s+t-u,s}
	    \\&\hspace{75mm}\times\hat{\tilde h}^{t,\tilde s}_u\Big(\negative\ceil t+\frac 32,r+\ceil t-\frac 32;\lambda\Big)\Bigg\}.
      \end{aligned}
   \end{gather}
Here we have used the relations given in equation \eqref{eq:StructureConstantPropertyForOPECalculation} and the step function defined as
  \begin{equation}
    \chi_{\mathcal A}(m) = \begin{cases}
			  1, \qquad m\in\mathcal A,\\
			  0, \qquad m\not\in\mathcal A.
			   \end{cases}
  \end{equation}
Note that the ``hat'' means we are using the structure constants of $\shs[\lambda]$, see appendix \ref{appendix:StructureConstantsOfHigherSpinAlgebra} for more details. Looking at the form of the equations given by $c^B_{s,m}$ and $c^F_{s,r}$ one can see that by starting from the highest modes, $m=\floor s-1$ and $r=\ceil s-\frac 32$, we can recursively solve $\mu^s_m$ and $\nu^s_r$ in terms of the highest modes $\mu^s_{\floor s-1}$ and $\nu^s_{\ceil s-\frac 32}$, respectively. Finally at the lowest modes, $m=-\floor s+1$ and $r=-\ceil s+\frac 32$, the equations of motion are reduced to relations containing only $\mathcal L_s$, $\psi_s$, $\mu^s_{\floor s-1}$ and $\nu^s_{\ceil s-\frac 32}$. These equations are the holographic Ward identities in the presence of sources \cite{Gutperle:2011kf,Banados:2004nr,Corley:2000ev}, and from these we can identify the correct normalization for the sources by holographically deriving the corresponding OPE's of the dual CFT.

Before we proceed, we will present a general result which will be very useful for us later.
\subsection{General Formula for Ward Identities from CFT}
One can derive a very useful and general formula for Ward identities in the presence of source terms. Consider two chiral quasi-primary fields $W(z)$ and $X(z)$ of conformal weights $h_W$ and $h_X$, respectively, and the following general OPE
  \begin{gather}
      \begin{split}
	W(z)X(w) &\sim \sum_{i=1}^{\infty}\frac{\sigma_i}{(z-w)^i}Z_i(w)
		  = \sum_{i=1}^{\infty}\frac{\sigma_i}{(i-1)!}\,\partial_w^{i-1}\left(\frac 1{z-w}\right)Z_i(w),
      \end{split}
  \end{gather}
where $Z_i(w)$ is are chiral quasi-primary fields of weight $h_i = h_W + h_X-i$ and we have used the compact notation $\sigma_iZ_i = \sum_j (\sigma_i)_j(Z_i)_j$ in case there are several fields with the same conformal weight. We are interested in expectation values of $W(z)$, but with insertions of $X(z)$ source terms
  \begin{equation}
   \average[\big]{W}_{\mu} = \average[\big]{W\,e^{-\int\mu X}},
  \end{equation}
where $\mu(w,\bar w)$ is a non-chiral source. Due to the insertion of $\mu(w,\bar w)$, the vacuum expectation value $\average[\big]{W}_\mu$ will gain $\bar z$ dependence. We can directly derive the following result
  \begin{gather}\label{eq:GeneralFormularForWardIdentitiesWithSources}
    \begin{aligned}
      \bar{\partial}&\average[\big]{W(z)}_{\mu} = - \bar{\partial}\,\average[\bigg]{\int\ud^2 w\,\mu(w,\bar w)\,
\contraction{}{W}{(z)}{X}W(z)X(w)}_{\mu}\\
					       &= 2\pi\,\average[\bigg]{\Big(\sigma_2\left[\partial Z_2\,\mu + Z_2\,\partial\mu\right]-\sigma_1\, Z_1\,\mu\Big) + \sum_{i=3}^{\infty}\frac{(-1)^i\,\sigma_i}{(i-1)!}\,\sum_{q=0}^{i-1}\begin{pmatrix}
										      i-1\\q
										    \end{pmatrix}
							  \partial^{i-1-q}Z_i\,\partial^q\mu}_{\mu},
    \end{aligned}
  \end{gather}
where we have used partial integration, the identity $\bar{\partial}\left(\frac 1{z-w}\right) = 2\pi\,\delta^{(2)}(z-w)$ and finally
  \begin{equation}
    \partial^n\big(Z\,\mu\big) = \sum_{q=0}^n\begin{pmatrix}
					      n\\q
                                             \end{pmatrix}
			  \partial^{n-q}Z\,\partial^q\mu.
  \end{equation}
For illustrative reasons, let us take two simple examples. Let $W=T$ be the energy-momentum tensor and $X$ a primary field, we then have the following data from their OPE $\sigma_1 = 1$, $\sigma_2=h_X$, $Z_1 = \partial X$, $Z_2 = X$ and all other coefficients are zero. This leads to the identity
  \begin{equation}
    \frac 1{2\pi}\bar{\partial}\average[\big]{T(z)}_{\mu_X} = \average[\big]{h_X\,X\,\partial\mu_X + (h_X-1)\,\partial X\mu_X}_{\mu_X}.
  \end{equation}
As a second example let us choose both fields to be the energy-momentum tensor $W = X = T$. For this case we have the following OPE coefficients $\sigma_1 = 1$, $\sigma_2=2$, $\sigma_4 = \frac c2$, $Z_1 = \partial T$, $Z_2 = T$ and $Z_4 = \mathbbm 1$. This leads to the following identity
  \begin{equation}
    \frac 1{2\pi}\bar{\partial}\average[\big]{T(z)}_{\mu_T} = \average[\big]{2T\,\partial\mu_T + \,\partial T\mu_T + \frac c{12}\,\partial^3\mu_T}_{\mu_T}.
  \end{equation}
As expected, this is just like the above result up to the central charge term. In the following we shall mainly use our result \eqref{eq:GeneralFormularForWardIdentitiesWithSources} the other way around, we will from the bulk derive the Ward identities then use \eqref{eq:GeneralFormularForWardIdentitiesWithSources} to find the OPE coefficients.
\subsection{Holographic Operator Product Expansions and Superconformal Symmetries}
The holomorphic conserved currents on the boundary can be organized into $\mathcal N=2$ multiplets
  \begin{equation}\label{eq:MultipletsDualCFT}
    \Big(W^{s-},G^{(s+\frac 12)-},G^{(s+\frac 12)+},W^{(s+1)+}\Big),\qquad s\in\mathbb Z_{\geq 1},
  \end{equation}
where $W^{s\pm}$ are bosonic fields of spin $s$ and and $G^{(s+\frac 12)\pm}$ are fermionic fields of spin $s+\frac 12$. The modes of these fields should form the $\mathcal N=2$ $\mathcal{SW}_{\infty}[\lambda]$ algebra, which generates the spectrum of the dual CFT. To begin with we will focus on the lowest multiplet $s=1$, which corresponds to the $\mathcal N=2$ superconformal Virasoro algebra. We use the notation $j\equiv W^{1-}$ and $T\equiv W^{2+}$ as is standard in the literature.

As discussed above, the terms with highest mode $\mu^s_{\floor s-1}$ and $\nu^s_{\ceil s-\frac 32}$ are the most dominant near the boundary and can thus be identified with sources of the dual fields up to normalization. So in order to find the Ward identities of this multiplet we only need to turn on boundary terms corresponding to these fields $\big(\mu^1_0,\nu^{\frac 32}_{\pm \frac 12}, \nu^2_{\pm\frac 12}, \mu^2_{\pm 1}\big)$, thus all other source terms are turned off. For reasons which will become more clear momentarily, we will rename $\mathcal L_2\rightarrow\tilde{\mathcal L}_2$. We can now recursively solve the equations \eqref{eq:EquationsToBeSolvedRecursivelyForTheSourceTermsAndThenToFindTheWardIdentities} in order to express all near-boundary terms in terms of the highest modes. The final equations for the lowest modes $c^B_{\frac 32, 0}=0$, $c^B_{2,-1}=0$, $c^F_{s, -\frac 12}=0$ (where $s=\frac 32, 2$) can be expressed in the following compact form
  \begin{gather}
    \begin{aligned}
    \bar{\partial}\mathcal L_{\frac 32} &= -\psi_2\,\nu^{\frac 32}_{\frac 12} - \psi_{\frac 32}\,\nu^2_{\frac 12} + \frac k{2\pi}\, 2\,\partial\mu^{\frac 32}_0,\\
    \bar{\partial}\tilde{\mathcal L}_2 &= 2\,\tilde{\mathcal L}_2\,\partial\mu^2_1 + \partial\tilde{\mathcal L}_2\,\mu^2_1 + \frac k{2\pi}\frac 12\,\partial^3\mu^2_1
		       +\sumCircles_{s=\frac 32}^2\bigg(\frac 32\psi_s\,\partial\nu^s_{\frac 12}+\frac 12\partial\psi_s\,\nu^s_{\frac 12}+\frac{2\pi}k\frac 12\,\mathcal L_{\frac 32}\,\psi_{\bar s}\,\nu^s_{\frac 12}\bigg),\\
    \bar{\partial}\psi_s &= \bigg(\frac 32\psi_s\,\partial\mu^2_1 + \partial\psi_s\,\mu^2_1 - \frac{2\pi}k\frac 12\,\mathcal L_{\frac 32}\,\psi_{\bar s}\,\mu^2_1\bigg) + \bigg(\psi_{\bar s}\,\mu^{\frac 32}_0\bigg) + (-1)^{2s}\bigg(\partial\mathcal L_{\frac 32}\,\nu^{\bar s}_{\frac 12} + 2\,\mathcal L_{\frac 32}\,\partial\nu^{\bar s}_{\frac 12}\bigg)\\
	 &\quad + (-1)^{2s}\bigg(\frac k{2\pi}\,2\,\partial^2\nu^s_{\frac 12}+2\,\tilde{\mathcal L}_2\,\nu^s_{\frac 12} + \frac{2\pi}k\frac 12\Big[\mathcal L_{\frac 32}\Big]^2\,\nu^s_{\frac 12}\bigg),
    \end{aligned}
  \end{gather}
where $\bar s=\frac 32$ if $s=2$ and $\bar s=2$ if $s=\frac 32$. These equations are the holographic Ward identities.
If we make the following identifications with the currents
  \begin{equation}
    2\pi\,\tilde{\mathcal L}_2\rightarrow \tilde T,\qquad 2\pi\,\mathcal L_{\frac 32}\rightarrow j,\qquad 2\pi\,\psi_{\frac 32}\rightarrow G^{\frac 32-},\qquad 2\pi\,\psi_2\rightarrow G^{\frac 32+},
  \end{equation}
and of boundary sources
  \begin{equation}\label{eq:SourceIdentifications}
    \mu^2_{-1}\rightarrow 2\pi\,\mu_{\tilde T},\qquad \mu^{\frac 32}_0\rightarrow 2\pi\,\mu_j,\qquad \nu^{\frac 32}_{-\frac 12}\rightarrow 2\pi\, \nu_{G^{\frac 32-}},\qquad \nu^2_{-\frac 12}\rightarrow 2\pi\, \nu_{G^{\frac 32+}},
  \end{equation}
we can use equation \eqref{eq:GeneralFormularForWardIdentitiesWithSources} to derive the following OPE coefficients of the dual currents. However it turns out that these OPE's are not the usual ones of $\mathcal N=2$ superconformal algebra. This can be fixed by the following Sugawara redefinition
  \begin{equation}
    T(z) = \tilde T(z) + \frac 1{4k}[jj](z).\label{eq:ShiftedEnergyMomentumTensorByRSymmetryAffineLieAlgebra}
  \end{equation}
Setting the Chern-Simons level to $k = \frac c6$, and we find the OPE's of the $\mathcal N=2$ superconformal algebra\footnote{Note that we are currently looking at the large $N$ limit of the duality, which means that the central charge is very large $c\rightarrow\infty$. In this ``classical'' limit we do not have any information about normal ordering, this means that we need to use ``classical'' OPE's. This means we can ignore double (and higher order) contractions when calculating OPE's, there are however $\mathcal O(\frac 1c)$ corrections when moving to finite $N$ due to quantum effects. See \cite{Gaberdiel:2012ku,Candu:2012tr} for some interesting analysis of the $\mathcal O(\frac 1c)$ corrections.}
  \begin{gather}
    \begin{aligned}
      j(z)j(w)&\sim \frac{c/3}{(z-w)^2},\qquad j(z)G^{\frac 32\pm}(w)\sim \frac 1{z-w}\,G^{\frac 32\mp}(w),\\
      T(z)T(w)&\sim \frac{c/2}{(z-w)^4} + \frac 2{(z-w)^2}\,T(w) + \frac 1{z-w}\,\partial T(w),\\
      T(z)G^{\frac 32\pm}(w)&\sim\frac{3/2}{(z-w)^2}\,G^{\frac 32\pm}(w) + \frac 1{z-w}\,\partial G^{\frac 32\pm}(w),\\
      G^{\frac 32\pm}(z)G^{\frac 32\pm}(w)&\sim \frac{\mp\,2c/3}{(z-w)^3}+\frac{\mp\, 2}{z-w}\,T(w),\\
      G^{\frac 32\pm}(z)G^{\frac 32\mp}(w)&\sim\frac{\pm\,2}{(z-w)^2}\,j(w) + \frac{\pm 1}{z-w}\,\partial j(w),\\
      T(z)j(w) &\sim \frac 1{(z-w)^2}\,j(w) + \frac 1{z-w}\,\partial j(w).
    \end{aligned}
  \end{gather}
Here $T(z)$ is the energy-momentum tensor and generates the Virasoro algebra, $j(z)$ is the $U(1)$ R-symmetry and generates an affine Lie algebra while $G^{\frac 32\pm}$ are the two conformal supercharges. In the literature the supercharges are chosen such that they have definite $U(1)$ charge under R-symmetry, this can be recovered from the superpositions $\tilde G^{\pm} = \frac i{\sqrt 2}(G^{\frac 32+} \pm G^{\frac 32-})$, for which
  \begin{gather}
    \begin{aligned}
    j(z)\tilde G^{\pm}(w)&\sim\frac{\pm 1}{z-w}\tilde G^{\pm}(w),\\
    \tilde G^\pm(z)\tilde G^\mp(w) &\sim \frac{2c/3}{(z-w)^3} \pm \frac 2{(z-w)^2}j(w) + \frac 1{z-w}\Big(2T(w)\pm \partial j(w)\Big),
    \end{aligned}
  \end{gather}
and finally $\tilde G^\pm(z)\tilde G^{\pm}(w)\sim 0$.
Note that combining $k=\frac l{4G}$ with $c=6k$, we find the celebrated Brown-Henneaux \cite{Brown:1986nw} central charge
  \begin{equation}
    c = \frac {3l}{2G}.
  \end{equation}
We note that the need for the Sugawara redefinition \eqref{eq:ShiftedEnergyMomentumTensorByRSymmetryAffineLieAlgebra} has been seen earlier in the literature \cite{Henneaux:1999ib, Hanaki:2012yf,Tan:2012xi}.

Let us now consider a general multiplet consisting of the fields \eqref{eq:MultipletsDualCFT}, turn off all source terms except the ones corresponding to this multiplet, use the Sugawara redefinition $\mathcal L_2 = \tilde{\mathcal L}_2 + \frac{\pi}{2k}[\mathcal L_{\frac 32}]^2$ and then proceed recursively.
Again identifying the currents and sources similar to equation \eqref{eq:SourceIdentifications} we find the following OPE's
  \begin{gather}
    \begin{aligned}
      T(z)W^{s-}(w) &\sim \frac s{(z-w)^2}\,W^{s-}(w) + \frac 1{z-w}\,\partial W^{s-}(w),\\
      T(z)W^{(s+1)+}(w) &\sim \frac{(s+1)}{(z-w)^2}\,W^{(s+1)+}(w) + \frac 1{z-w}\,\partial W^{(s+1)+}(w),\\
      T(z)G^{(s+\frac 12)\pm}(w) &\sim \frac {s+1/2}{(z-w)^2}\,G^{(s+\frac 12)\pm}(w) + \frac 1{z-w}\,\partial G^{(s+\frac 12)\pm}(w),\\
      j(z)G^{(s+\frac 12)\pm}&\sim \frac 1{z-w}G^{(s+\frac 12)\mp}(w).
    \end{aligned}
  \end{gather}
This we have checked for many low spins $s$. By the exact same procedure it is possible to derive OPE's between higher-spin fields and thereby the structure constants of the (classical) non-linear $\mathcal N=2$ $\mathcal{SW}_{\infty}[\lambda]$ algebra. Note that beside the Sugawara redefinition of the Energy-Momentum tensor, it is not possible to make any non-linear redefinitions that respect the $\mathcal N=2$ multiplets. A systematic analysis of the OPE's between the higher-spin field is outside the scope of this paper.

Let us however show one last general result which will be useful to us later, that the leading-order singularity of the OPE of higher-spin bosonic currents is found from the term
  \begin{equation}\label{eq:HolographicWardIdentityLeadingTermOfHigherSpinOPEs}
    \bar{\partial}\mathcal L_s = -\frac{\frac k{2\pi}\,N^B_s}{\big(2\floor s-2\big)!}\,(-\partial)^{2\floor s-1}\mu^s_{\floor s-1} + \dots,
  \end{equation}
which leads to the following leading-order term
  \begin{equation}
    W_s(z)W_s(w) \sim \frac{-k\,N^B_s\,\big(2\floor s-1\big)}{(z-w)^{2\floor s}} + \dots,
  \end{equation}
where for simplicity we use the notation that for integer $s\in\mathbb Z$ we have the fields $W_s = W^{s+}$, while for half-integers $s=\floor s+\frac 12\in\mathbb Z+\frac 12$ we have $W_{\floor s+\frac 12} = W^{\floor s -}$. Note that the leading-order term of $W^{2-}W^{2-}$ exactly matches the results of \cite{Hanaki:2012yf}, up to a sign due to differing normalizations.

In this section we have established the precise AdS/CFT dictionary for the higher-spin fields. We have in particular shown that using the normalizations given in \eqref{eq:NormalizationOfSourceTermsUsingTrace}, we can identify the bulk terms $\frac 1{2\pi} \mu^s_{\floor s-1}$ and $\frac 1{2\pi} \nu^s_{\ceil s-\frac 32}$ with source terms of the boundary CFT \eqref{eq:BoundaryActionAddedSourceTerm}.\footnote{This would seem to imply that the factors of $1/(2\pi)$ in the three-point functions of \cite{Ammon:2011ua} ought to be absent.}
\section{Three-point functions from the bulk} \label{Bulk}
We have so far found that our formalism reproduces the correct masses of the scalars in Vasiliev theory and established which terms in the bulk gauge connection correspond to source terms of which dual higher-spin current, and along the way given an alternative proof of the emergence of superconformal $\mathcal N=2$ $\mathcal{SW}_{\infty}[\lambda]$ symmetry near the $\AdS_3$ boundary. In this section we will use this information to calculate certain classes of three-point functions containing two scalars and one (holomorphic) bosonic higher-spin current. This section follows closely the ideas developed in \cite{Ammon:2011ua}.

For our needs we can turn off all higher-spin fields in the bulk, except one of fixed spin $s$. The gauge connection will take the form
  \begin{equation}\label{eq:GaugeConnectionWithOnlyOneSpinAndSourceTermTurnedOn}
    A = \left(e^{\rho}\, L^{(2)}_1 + \frac 1{B_s}\,e^{-(\floor s-1)\rho}\mathcal L_s\,L^{(s)}_{-\floor s+1}\right)\,\ud z + \sum_{|m|\leq\floor s -1}e^{m\rho}\,\mu^s_m\, L^{(s)}_m\,\ud\bar z + L_0\,\ud\rho,
  \end{equation}
where out of convenience we will in the following use the notation 
  \begin{equation}
    \frac 1{B_s} \equiv \frac {2\pi}k\frac 1{N^B_s},\qquad \frac 1{F_s} \equiv \frac{2\pi}k\frac 1{N^F_s}.
  \end{equation}

Using the standard methods of the AdS/CFT correspondence to calculate correlation functions is too cumbersome and does not take full advantage of the higher-spin gauge symmetries. Our strategy for calculating three-point functions of the form $\average[\big]{\mathcal O_{\Delta}(z_1,\bar z_1)\bar{\mathcal O}_{\Delta}(z_2,\bar z_2) J^s(z_3)}$ is based on the observation made in \cite{Ammon:2011ua}. Starting from the solution of a free scalar field on $\AdS_3$ we can generate new solutions by performing higher-spin gauge transformations\footnote{The gauge transformations we are using are non-vanishing at the boundary and therefore are not real gauge transformations. In other words, they act like global symmetries since they map a configuration to a physically distinct one.}. We can therefore start from scalars on $\AdS_3$, then by a gauge transformation introduce higher-spin source terms. From the near boundary expansion of the scalars we can then find the corresponding three-point functions. This means we can reduce the whole calculation into studying how the scalars transform under higher-spin gauge symmetries.

The gauge transformation which maps the $\AdS_3$ connection into a chiral higher-spin background with spin $s$ and its boundary source term \eqref{eq:GaugeConnectionWithOnlyOneSpinAndSourceTermTurnedOn} is of the following form
  \begin{gather}
    \begin{aligned}
      \Lambda(\rho,z,\bar z) = \sum_{m=0}^{\floor s-1}\frac 1{\big(\floor s-m-1\big)!}\big(\negative\partial\big)^{\floor s-m-1}\Lambda^s\, e^{m\rho}\,L^{(s)}_m + \sum_{m=0}^{\floor s-1}\tilde F^s_{-m}\, e^{-m\rho}\,L^{(s)}_{-m}
    \end{aligned}
  \end{gather}
and $\bar{\Lambda}(\rho,z,\bar z) = 0$, with the following identifications: $\mu^s_{\floor s-1} = \bar{\partial}\Lambda$ and $\mathcal L_s = -\frac{B_s}{\left(2\floor s-2\right)!}(\negative\partial)^{2\floor s-1}\Lambda$, which is imposed by the equations of motion \eqref{eq:HolographicWardIdentityLeadingTermOfHigherSpinOPEs}. Note that $\tilde F^s_{-m}$ can be explicitly found by using the equations of motion. But as we will briefly see, the negative mode contributions to the connection do not contribute to the three-point functions.

Under infinitesimal gauge transformations, the matter fields transform as
  \begin{equation}
    \widehat C = C + \delta_s C,\qquad \delta_s C = C\star\bar\Lambda - \Lambda\star C = -\Lambda\star C.
  \end{equation}
Putting the fermions $C^s_r$ to zero in \eqref{eq:ExpansionOfGaugeAndMatterFieldInTermsOfSuperHigherSpinAlgebra} we find that the generating function transforms as
  \begin{gather}
    \begin{aligned}
      \delta_s C &= -\sumCircles_{t=1}^{\infty}\sum_{|n|\leq\floor t-1}\sum_{m=0}^{\floor s-1}\frac{\big(\negative\partial\big)^{\floor s-m-1}\Lambda^s}{\big(\floor s-m-1\big)!}\, C^t_n\, e^{m\rho}\,L^{(s)}_m\star L^{(t)}_n + \underbrace{\ldots}_{m<0}\\
	       &= \delta_s C^1_0\, L^{(1)}_0 + \delta_s C^{\frac 32}_0\, L^{(\frac 32)}_0 + \ldots
    \end{aligned}
  \end{gather}
In order to isolate how the scalars transform, recall that
    \[
	L^{(s)}_m\star L^{(t)}_n = \sumCircles_{u=1}^{\text{Min}(2s-1,2t-1)}g^{st}_u(m,n;\lambda)\, L^{(s+t-u)}_{m+n}.
    \]
In order to isolate the lowest two scalars, we have the following conditions
  \begin{gather*}
    \begin{aligned}
      m + n &= 0	\quad	&\Rightarrow	\quad	m&=-n,\\
      s+t-u_q &=q	\quad	&\Rightarrow	\quad	u_q &= s+t-q,
    \end{aligned}
  \end{gather*}
where $q=1,\frac 32$. Now for $q=1$, if $t>s$ or $s>t$ we have that $u_1>\text{Min}(2s-1,2t-1)$ which implies that $g^{st}_{u_q}(\dots) = 0$. This implies that only the term with $s=t$ contributes. For $q=\frac 32$, besides the $t=s$ terms also the $t = s\pm\frac 12$ terms contribute. Thus the scalars transform as
  \begin{equation}
    \delta_s C^1_0 = -\sum_{m=0}^{\floor s-1}\frac{\big(\negative\partial\big)^{\floor s-m-1}\Lambda^s}{\big(\floor s-m-1\big)!}\, C^s_{-m}\, g^{ss}_{2s-1}\big(m,\negative m;\lambda\big)\, e^{m\rho} +
		      \begin{aligned}
			&\text{\footnotesize terms which}\\&\text{\footnotesize vanish as $\rho\rightarrow\infty$}
		      \end{aligned},
  \end{equation}
and
    \begin{align}
      \delta_s C^{\frac 32}_0 &= -\sum_{m=0}^{\floor s-1}\frac{\big(\negative\partial\big)^{\floor s-m-1}\Lambda^s}{\big(\floor s-m-1\big)!}\,  \bigg[C^s_{-m}\,g^{ss}_{2s-\frac 32}\big(m,\negative m;\lambda\big)
	\\ &\quad +C^{s-1/2}_{-m}\,g^{ss-1/2}_{2s-2}\big(m,\negative m;\lambda\big)\,\chi_{\big[0,\floor{s-1/2}-1\big]}(m)+C^{s+1/2}_{-m}\,g^{ss+1/2}_{2s-1}\big(m,\negative m;\lambda\big)\bigg]\, e^{m\rho}.\nonumber
     \end{align}
The step function in the second term is put in to ensure we do not go beyond the wedge algebra. Using this we can readily find the transformation of the mass-eigenstates $\widehat{\phi}_i = \phi_i +\delta\phi_i$
  \begin{gather} \label{eq:InfinitesimalGaugeTransformationOfMassEigenStates}
    \begin{aligned}
      \delta_s\phi_i &= \tilde a_i\,\delta_s C^1_0 + \tilde b_i\,\delta_s C^{\frac 32}_0,\\
		   &= -\sum_{m=0}^{\floor s-1}\frac{\big(\negative\partial\big)^{\floor s-m-1}\Lambda^s}{\big(\floor s-m-1\big)!}\, e^{m\rho}\,\Bigg(\tilde a_i\,C^s_{-m}\,g^{ss}_{2s-1}\big(m,\negative m;\lambda\big) + \tilde b_i\,\bigg[C^s_{-m}\,g^{ss}_{2s-\frac 32}\big(m,\negative m;\lambda\big)
	 \\&\quad+C^{s-1/2}_{-m}\,g^{ss-1/2}_{2s-2}\big(m,\negative m;\lambda\big)\chi_{\big[0,\floor{s-1/2}-1\big]}(m)+C^{s+1/2}_{-m}\,g^{ss+1/2}_{2s-1}\big(m,\negative m;\lambda\big)\bigg]\Bigg),\\
%%%
		   &\equiv \sum_{m=0}^{\floor s-1}\Big[f^{s,i}_m\big(\lambda,\partial_{\rho}\big)\,\partial^m\phi_i\Big]\,\partial^{\floor s-m-1}\Lambda^s,\\
		   &\equiv D^{(s,i)}(z)\phi_i.
    \end{aligned}
\end{gather}
This expression requires solving the recursion relations \eqref{eq:VasilievEquationRecursionRelationsForScalars} in order to express the auxiliary fields $C^s_{-m}$ as sums and derivatives of $C^1_0$ and $C^{\frac 32}_0$, which in turn can be expressed as functions of $\phi_{\pm}$. As seen later, it turns out that these will have the form $C^s_{-m}\sim e^{-|m|\rho}\,A\big(\lambda,\partial_{\rho}\big)\,\partial^m\phi_i$,\footnote{Note that for our calculation of three-point functions we only need to turn on the boundary source of the relevant scalar. Thus in calculating $\delta_s\phi_+$ we set $\phi_-=0$ and vice versa.} which means that $e^{m\rho}$ is canceled for $m>0$ and enhanced for $m<0$. For this reason the terms with $m<0$ has been neglected in \eqref{eq:InfinitesimalGaugeTransformationOfMassEigenStates}, since they are vanishing near the $\AdS_3$ boundary. The coefficients are given as
  \begin{gather}
    \tilde a_i = \begin{cases}
		  -1,\quad &i=+,\\
		  \;\;\;1,\quad &i=-,
                 \end{cases},
    \qquad
    \tilde b_i = \begin{cases}
		  \;\;\;2\lambda,\quad &i=+\\
		  -2\lambda+1\quad &i=-
                 \end{cases},
  \end{gather}
which are found by inverting the equations \eqref{eq:VasilievLowestComponentScalarAsSuperpositionsOfPhysicalScalars}. The function in the third line of \eqref{eq:InfinitesimalGaugeTransformationOfMassEigenStates} contains all the information about the higher-spin deformation and is given as 
  \begin{gather}\label{eq:HigherSpinDeformationFunctionFDefinitnion}
    \begin{align}
      f^{s,i}_m\big(\lambda,\partial_{\rho}\big) &= \frac{(-1)^{\floor s-m}}{\big(\floor s-m-1\big)!}\,\Bigg(\tilde a_i\,\mathcal G^{s,i}_m\,g^{ss}_{2s-1}\big(m,\negative m;\lambda\big) + \tilde b_i\,\bigg[\mathcal G^{s,i}_m\,g^{ss}_{2s-\frac 32}\big(m,\negative m;\lambda\big)
				\\&\quad+\mathcal G^{s-1/2,i}_m\,g^{ss-1/2}_{2s-2}\big(m,\negative m;\lambda\big)\chi_{\big[0,\floor{s-1/2}-1\big]}(m)+\mathcal G^{s+1/2,i}_m\,g^{ss+1/2}_{2s-1}\big(m,\negative m;\lambda\big)\bigg]\Bigg),\nonumber
    \end{align}
  \end{gather}
where $\mathcal G^{s,i}_m$ is defined as
    \[
      e^{-|m|\rho}\,\mathcal G^{s,i}_m\big(\lambda,\partial_{\rho}\big)\,\partial^m\phi_i = C^s_{-m}(\lambda,\partial_{\rho})\big|_{\phi_{\bar i}=0},
    \]
where $i = \pm$ and the index $\bar i$ refers to the opposite sign. Thus we find $\mathcal G^{s,i}_m$ by removing a factor of $e^{-|m|\rho}\,\partial^m\phi_i$ from $C^s_{-m}$ and setting the other scalar to zero.

\subsection{Three-Point Functions}
Recall that putting a scalar source on the boundary of $\AdS_3$ at $z'$, we can express the bulk solution using the bulk-to-boundary propagator
  \begin{equation}\label{eq:ScalarOnAdS3UsingBulkToBoundaryPropagator}
    \phi_i(\rho,z) = \int\ud^2z' \,G_{b\partial}(\rho,z;z')\,\phi_i^{\partial}(z'),
  \end{equation}
which in our coordinates is given as \cite{Witten:1998qj,Freedman:1998tz}
  \begin{equation}\label{eq:BulkToBoundaryPropagatorOfAdS3}
    G_{b\partial}(\rho,z;z') = c_{\pm}\,\left(\frac{e^{-\rho}}{e^{-2\rho} + |z-z'|^2}\right)^{\Delta_{\pm}},
  \end{equation}
where $c_{\pm} =  \frac{\Gamma(\Delta_{\pm})}{\pi\,\Gamma(\Delta_{\pm}-1)} = \frac{\Delta_{\pm}-1}{\pi}$.
Here the conformal weights are determined from the scalar masses $m^2 = \Delta_{\pm}(\Delta_{\pm}-2)$, where $\Delta_+\geq\Delta_-$ and $\Delta_{\pm} = 2-\Delta_{\mp}$.
In this section we will also use the conventional coordinates $r = e^{-\rho}$, in which the metric takes the form $\ud s^2 =\frac{\ud r^2+\ud z\ud\bar z}{r^2}$ and the boundary is at $r\rightarrow 0$.
The near-boundary expansion of the bulk field is of the form \cite{Klebanov:1999tb}
  \begin{equation}
    \phi_i(\rho,z)\longrightarrow r^{d-\Delta_{\pm}}\left(\phi^{\partial}_i(z) + o(r)\right) + r^{\Delta_{\pm}}\left(\frac 1{B_{\phi}^{\pm}}\,\average[\big]{\mathcal O_{\Delta_{\pm}}(z)} + o(r)\right),
  \end{equation}
where $\mathcal O_{\Delta_{\pm}}$ is the dual field with conformal weight $\Delta_{\pm}$ and $B_{\phi}^{\pm} = 2\Delta_{\pm}-d$ is necessary for a consistent dictionary \cite{Freedman:1998tz,Klebanov:1999tb}. The idea is to generate the solution on a background containing a spin $s$ source by a gauge transformation
  \begin{gather}\label{eq:ScalarFieldInfinitesimalGaugeTransformed}
    \begin{split}
      \phi_i(\rho, z)\longrightarrow \widehat{\phi}_i(\rho,z) &= \phi_i(\rho,z) + \delta_s\phi_i(\rho,z),\\
						      &= \big(1 + D^{(s,i)}\big)\,\phi_i(\rho,z),
    \end{split}
  \end{gather}
which gives the near boundary expansion
  \begin{equation}\label{eq:NearBoundaryExpansionOfTheGaugedTransformedScalarField}
    \widehat{\phi}_i(\rho,z)\longrightarrow r^{d-\Delta_{\pm}}\left(\widehat{\phi}^{\partial}_i(z) + o(r)\right) + r^{\Delta_{\pm}}\left(\frac 1{B_{\phi}^{\pm}}\,\average[\big]{\mathcal O_{\Delta_{\pm}}(z)}_{\mu} + o(r)\right).
  \end{equation}
The notation $\average{\dots}_{\mu}$ stands for the vacuum expectation value, with a higher-spin source insertion. We will put a scalar point-source at $z_2$ and a chiral spin $s$ source at $z_3$ on the $\AdS_3$ boundary
   \begin{equation}\label{eq:BoundaryConditionsSorcesForScalarAndSpinSField}
      \widehat\phi_i^{\partial}(z,\bar z) = \mu_{\phi}\,\delta^{(2)}(z-z_2),\qquad \mu^s_{\floor s-1}(z,\bar z) = \mu_s\,\delta^{(2)}(z-z_3).
   \end{equation}
The two and three-point functions can then be read off from the one-point function near the boundary
  \begin{gather}\label{eq:OnePointFunctionWithInsertionsSources}
    \begin{aligned}
    \average[\big]{\mathcal O_{\Delta_{\pm}}(z_1,\bar z_1)}_{\mu} = &\mu_{\phi}\,\average[\big]{\mathcal O_{\Delta_{\pm}}(z_1,\bar z_1)\,\bar{\mathcal O}_{\Delta_{\pm}}(z_2,\bar z_2)}
	     \\&\quad + \mu_{\phi}\,\mu_s\,\average[\big]{\mathcal O_{\Delta_{\pm}}(z_1,\bar z_1)\,\bar{\mathcal O}_{\Delta_{\pm}}(z_2,\bar z_2)\, J^{s}(z_3)} + \ldots.
    \end{aligned}
  \end{gather}
We will now find a general expression for the three-point functions as a function of $f^{s,i}_m(\lambda,\partial_\rho)$ given in equation \eqref{eq:HigherSpinDeformationFunctionFDefinitnion}, which characterize the higher-spin deformation the scalars experience. The steps are clear; we need to write down how the scalars transform \eqref{eq:ScalarFieldInfinitesimalGaugeTransformed} and use \eqref{eq:ScalarOnAdS3UsingBulkToBoundaryPropagator}, which requires knowing $\phi_i^{\partial}$ as a function of $\widehat\phi_i^{\partial}$. Next we need to find the vacuum expectation value of the dual field from the asymptotics of $\hat\phi_i$ \eqref{eq:NearBoundaryExpansionOfTheGaugedTransformedScalarField}, then isolate the $\mu_{\phi}\,\mu_s$ order contribution, which gives us the three-point functions as seen in \eqref{eq:OnePointFunctionWithInsertionsSources}.

The boundary sources of $\phi_i$ and $\widehat\phi_i$ are related by a gauge transformation
  \[	\widehat\phi_i^{\partial}(z)\,e^{-\Delta_{\mp}\rho} = \big(1+D^{(s,i)}\big)e^{-\Delta_{\mp}\rho}\,\phi_i^{\partial}(z) = e^{-\Delta_{\mp}\rho}\,\big(1+D_{\mp}^{(s,i)}\big)\,\phi_i^{\partial}(z),	\]
where we have defined
  \begin{equation}
    D^{(s,i)}_{\pm} = D^{(s,i)}\big(\partial_{\rho}\rightarrow -\Delta_{\pm}\big).
  \end{equation}
Inverting this up to first order and using the boundary condition \eqref{eq:BoundaryConditionsSorcesForScalarAndSpinSField} we find
  \begin{equation}
    \phi_i^{\partial}(z,\bar z) = \mu_{\phi}\,\big(1-D^{(s,i)}_{\mp}\big)\,\delta^{(2)}(z-z_2).
  \end{equation}
Using this, the gauge transformed scalar field is
  \begin{equation}
    \hat\phi(\rho,z) = \mu_{\phi}\,\big(1+D^{(s,i)}(z)\big)\,\int\ud^2z'\,G_{b\partial}(\rho,z;z')\,\big(1-D^{(s,i)}_{\mp}(z')\big)\,\delta^{(2)}(z'-z_2).
  \end{equation}
Going near the boundary $\rho\rightarrow\infty$ and keeping only the $e^{-\Delta_{\pm}\rho}$ contribution we have
  \begin{gather}
    \begin{aligned}\label{eq:ScalarPhiGaugeTransformedInOrderToFindTwoAndThreePointFunctions}
      \hat\phi_i(\rho,z) &\approx \mu_{\phi}\,\Big(1+D^{(s,i)}(z)\Big)\,\int\ud^2z'\,\frac{c_{\pm}\,e^{-\Delta_{\pm}\rho}}{|z-z'|^{2\Delta_{\pm}}}\,\Big(1-D^{(s,i)}_{\mp}(z')\Big)\,\delta^{(2)}(z'-z_2),\\
			 &= e^{-\Delta_{\pm}\rho}\,\mu_{\phi}\,c_{\pm}\,\int\ud^2 z'\,\Big(1+ D_{\pm}^{(i,s)}(z)\Big)\, \frac 1{|z-z'|^{2\Delta_{\pm}}}\,\Big(1-D_{\mp}^{(s,i)}(z')\Big)\,\delta^{(2)}(z'-z_2),\\
			 &=e^{-\Delta_{\pm}\rho}\,\frac{\average[\big]{\mathcal O(z)}_{\mu}}{B_{\phi}^{\pm}}, \qquad \rho\rightarrow\infty.
    \end{aligned}
  \end{gather}
The two-point function is readily given as
  \begin{equation}
    \average[\big]{\mathcal O_{\Delta_{\pm}}(z_1,\bar z_1)\,\bar{\mathcal O}_{\Delta_{\pm}}(z_2,\bar z_2)} = \frac{B_{\phi}^{\pm}\, c_{\pm}}{|z_1-z_2|^{2\Delta_{\pm}}}.
  \end{equation}
Next we will look at the $\mu_{\phi}\,D^{(s,i)}$ contribution of the one-point function given in \eqref{eq:ScalarPhiGaugeTransformedInOrderToFindTwoAndThreePointFunctions}, since $D^{(s,i)}$ is proportional to $\mu_s$. Neglecting the other terms, we have
  \begin{gather}\label{eq:OnePointFunctionButOnlyWithSourcesForThreePointFunction}
    \average[\big]{\mathcal O_{\Delta_{\pm}}(z_1)}_{\mu} = \mu_{\phi}\, B_{\phi}^{\pm}\, c_{\pm}\, \left[D^{(s,i)}_{\pm}(z_1)\,\frac 1{|z_1-z_2|^{2\Delta_{\pm}}}-\int\ud^2 z'\frac{D_{\mp}^{(s,i)}(z')\,\delta^{(2)}(z'-z_2)}{|z_1-z'|^{2\Delta_{\pm}}}\right].
  \end{gather}
Recall that the differential operators describing the infinitesimal gauge transformations take the form
  \begin{equation}
    D^{(s,i)}_\pm(z) = \sum_ {m=0}^{\floor s-1}\left[f^{s,i}_m\big(\lambda,-\Delta_{\pm}\big)\,\partial^{\floor s-m-1}\Lambda^s\right]\partial^m + \text{\footnotesize{terms vanishing as $\rho\rightarrow\infty$}}.
  \end{equation}
Using this and the following identity
  \begin{equation}\label{eq:FormulaForDifferentiationOfAbsoluteValueOfzStuffThingelyDingly}
    \partial_2^n\frac 1{|z_1-z_2|^{2\Delta_\pm}} = (-1)^n\partial_1^n\frac 1{|z_1-z_2|^{2\Delta_{\pm}}} = \frac{\Gamma(\Delta_\pm +n)}{\Gamma(\Delta_\pm)}\frac 1{(z_1-z_2)^n}\frac 1{|z_1-z_2|^{2\Delta_\pm}},
  \end{equation}
we can write the first term of \eqref{eq:OnePointFunctionButOnlyWithSourcesForThreePointFunction} as
  \begin{equation}
    D_\pm^{(s,i)}(z_1)\frac 1{|z_{12}|^{2\Delta_\pm}} = \sum_{m=0}^{\floor s-1}(-1)^m\frac{\Gamma(\Delta_\pm+m)}{\Gamma(\Delta_\pm)}\, f^{s,i}_m\big(\lambda,-\Delta_\pm\big)\left[\partial_1^{\floor s-m-1}\Lambda^{(s)}(z_1)\right]\frac 1{z_{12}^m\,|z_{12}|^{2\Delta_\pm}}.
  \end{equation}
For the second term we need to integrate by parts, until there are no derivatives on the delta function
%%  \begin{gather}
    \begin{align*}
      \int\ud^2\,z'&\frac{D_{\mp}^{(s,i)}(z')\,\delta^{(2)}(z'-z_2)}{|z_1-z'|^{2\Delta_{\pm}}} = \sum_{m=0}^{\floor s-1}f^{s,i}_m\big(\lambda,-\Delta_\mp\big)\int\ud^2z'\,\frac{\partial_{z'}^{\floor s-m-1}\Lambda^{(s)}(z')\,\partial_{z'}^m\delta(z'-z_2)}{|z_1-z'|^{2\Delta_\pm}},\\
	 &=\sum_{m=0}^{\floor s-1}f^{s,i}_m\big(\lambda,-\Delta_\mp\big)\int\ud^2z'\,(-1)^m\,\partial^m_{z'}\left[\frac{\partial_{z'}^{\floor s-m-1}\Lambda^{(s)}(z')}{|z_1-z'|^{2\Delta_\pm}}\right]\delta(z'-z_2),\\
	 &=\sum_{m=0}^{\floor s-1}(-1)^m\,f^{s,i}_m\big(\lambda,-\Delta_\mp\big)\,\partial_2^m\left[\frac{\partial_2^{\floor s-m-1}\Lambda^{(s)}(z_2)}{|z_{12}|^{2\Delta_\pm}}\right],\\
	 &=\sum_{m=0}^{\floor s-1}(-1)^m\, f^{s,i}_m\big(\lambda,-\Delta_\mp\big)\sum_{j=0}^m\begin{pmatrix}
											      m\\j
	                                                                                     \end{pmatrix}
	      \left[\partial_2^{\floor s-m-1+j}\Lambda^{(s)}(z_2)\right]\,\partial_2^{m-j}\left[\frac 1{|z_{12}|^{2\Delta_\pm}}\right],\\
	 &=\sum_{m=0}^{\floor s-1}\sum_{j=0}^m (-1)^m\frac{\Gamma(\Delta_\pm+m-j)}{\Gamma(\Delta_\pm)}\, f^{s,i}_m(\lambda,-\Delta_\mp)\,\begin{pmatrix}
																	    m\\j
	                                                                                                                                 \end{pmatrix}
		    \left[\partial_2^{\floor s-m-1+j}\Lambda^{(s)}(z_2)\right]\,\frac 1{z_{12}^{m-j}\,|z_{12}|^{2\Delta_\pm}},
    \end{align*}
%%  \end{gather}
where in the last line we have used the formula \eqref{eq:FormulaForDifferentiationOfAbsoluteValueOfzStuffThingelyDingly}. In order to get the correct boundary condition for the higher-spin field \eqref{eq:BoundaryConditionsSorcesForScalarAndSpinSField}, we have to set the gauge parameter to
  \begin{equation}
    \Lambda^{(s)}(z) = \frac{\mu_s}{2\pi}\,\frac 1{z-z_3}.
  \end{equation}
We can now make use of the identities
  \begin{gather}
    \begin{split}
      \partial_1^{\floor s-m-1}\Lambda^{(s)}(z_1) &= \frac{\mu_s}{2\pi}\,\frac{\big(\floor s-m-1\big)!}{z_{13}^{\floor s-m}}(-1)^{\floor s-m-1},\\
      \partial_2^{\floor s-m-1+j}\Lambda^{(s)}(z_2) &= \frac{\mu_s}{2\pi}\,\frac{\big(\floor s-m-1+j\big)!}{z_{23}^{\floor s-m+j}}(-1)^{\floor s-m-1+j},
    \end{split}
  \end{gather}
and write the one-point function as
  \begin{gather}
    \begin{aligned}\label{eq:FullExpressionOfThreePointFunctionNumberOne}
      \average[\big]{\mathcal O_{\Delta_\pm}&(z_1)}_{\mu} = \frac{\mu_{\phi}\,\mu_s\, B^\pm_m\,c_\pm\,(-1)^{\floor s-1}}{2\pi\, |z_{12}|^{2\Delta_\pm}}\sum_{m=0}^{\floor s-1}\frac 1{z^m_{12}}\,\Bigg\{f^{s,i}_s\big(\lambda,-\Delta_\pm\big)\,\frac{\Gamma(\Delta_\pm+m)}{\Gamma(\Delta_\pm)}\,\frac{\big(\floor s-m-1\big)!}{z_{13}^{\floor s-m}}\\
	  &- f^{s,i}_m\big(\lambda,-\Delta_\mp\big)\frac 1{z_{23}^{\floor s-m}}\sum_{j=0}^m(-1)^j\begin{pmatrix}
	                                                                                                m \\ j
	                                                                                               \end{pmatrix}
	      \frac{\Gamma(\Delta_\pm+m-j)}{\Gamma(\Delta_\pm)}\big(\floor s-m-1+j\big)!\left(\frac{z_{12}}{z_{23}}\right)^j \Bigg\}
    \end{aligned}
  \end{gather}
We have now shown that the three-point functions are known as soon as we know the functions $f^{s,i}_m(\lambda,\Delta_\pm)$. This expression, however, looks very complicated and it is not manifestly conformal invariant. Conformal symmetry constrains the three-point functions to take the form \footnote{Note that in general $z^{2h}\bar z^{2\bar h} = |z|^{2\Delta}e^{i(h-\bar h)\theta}$. For scalars we have that $h-\bar h =0$, while for spin $\frac 12$ fermions we have $h-\bar h = \pm\frac 12$.}
  \begin{gather}\label{eq:ThreePointFunctionzDependenceRequiredByConformalSymmetry}
      \begin{aligned}
	\average[\big]{\mathcal O_{\Delta_\pm}(z_1,\bar z_1)\bar{\mathcal O}_{\Delta_\pm}(z_2,\bar z_2)J^{(s)}(z_3)} &= A_\pm(s)\,d_\pm\,\left(\frac{z_{12}}{z_{13}z_{23}}\right)^{\floor s}\,\frac 1{z_{12}^{2h_\pm}\bar z_{12}^{2\bar h_\pm}},\\
				  &= A_\pm(s)\, \left(\frac{z_{12}}{z_{13}z_{23}}\right)^{\floor s}\,\average[\big]{\mathcal O_{\Delta_\pm}(z_1,\bar z_1)\bar{\mathcal O}_{\Delta_\pm}(z_2,\bar z_2)}.
      \end{aligned}
  \end{gather}
Note that this, among other things, demands the following relation
  \begin{equation}\label{eq:ThreePointFunctionSymmetryWhenExchangingTheFirstTwoCoordinates}
    	\average[\big]{\mathcal O_{\Delta_\pm}(z_1,\bar z_1)\bar{\mathcal O}_{\Delta_\pm}(z_2,\bar z_2)J^{(s)}(z_3)} = 	(-1)^{\floor s}\average[\big]{\mathcal O_{\Delta_\pm}(z_2,\bar z_2)\bar{\mathcal O}_{\Delta_\pm}(z_1,\bar z_1)J^{(s)}(z_3)}.
  \end{equation}
Although the full conformal invariance is not manifest in \eqref{eq:FullExpressionOfThreePointFunctionNumberOne}, we can make the above symmetry manifest in order to simplify \eqref{eq:FullExpressionOfThreePointFunctionNumberOne}. This implies that the three-point function must be of the form
  \begin{gather}\label{eq:FullExpressionOfThreePointFunctionNumberTwo}
    \begin{aligned}
      \average[\big]{\mathcal O_{\Delta_\pm}(z)}_{\mu} &= \mu_{\phi}\,B_{\phi}^{\pm}\, C_{\pm}\,\Big[D_{\pm}^{(s,i)}(z_1) + (-1)^{\floor s}\, D_{\pm}^{(s,i)}(z_2)\Big]\, \frac 1{|z_{12}|^{2\Delta_{\pm}}},\\
			 &= \frac{\mu_{\phi}\,\mu_s\,B^\pm_\phi\,C_{\pm}\,(-1)^{\floor s-1}}{2\pi\,|z_{12}|^{2\,\Delta_{\pm}}}\sum_{m=0}^{\floor s-1}\frac{f^{s,i}_m(\lambda,-\Delta_{\pm})}{z_{12}^m}\frac{\Gamma(\Delta_{\pm} + m)}{\Gamma(\Delta_{\pm})}\,\big(\floor s-m-1\big)!\\
			  &\quad\times\left(\frac 1{z_{13}^{\floor s-m}}+ \frac{(- 1)^{\floor s-m}}{z_{23}^{\floor s-m}}\right).
    \end{aligned}
  \end{gather}
Note that the second term is acting on $z_2$, thus the factor $(-1)^{\floor s-m}$ comes from using the formula \eqref{eq:FormulaForDifferentiationOfAbsoluteValueOfzStuffThingelyDingly}. Furthermore note that making the symmetry \eqref{eq:ThreePointFunctionSymmetryWhenExchangingTheFirstTwoCoordinates} manifest imposes a constraint on $f^{s,i}_m(\lambda,\Delta_\pm)$, which comes from equating \eqref{eq:FullExpressionOfThreePointFunctionNumberOne} with \eqref{eq:FullExpressionOfThreePointFunctionNumberTwo} and isolating terms of equal powers of $z_{12}$
  \begin{equation}\label{eq:ConstraintsOnThefFunction}
    f^{s,i}_{\floor s-\tilde j}\big(\lambda,-\Delta_\pm\big) = -\sum_{m=0}^{\floor s-1}(-1)^{\floor s-m}\,f^{s,i}_m\big(\lambda,-\Delta_\mp\big)\begin{pmatrix}
																		m\\ \tilde j-\floor s+m
                                                                                                                                              \end{pmatrix},
  \end{equation}
where $\tilde j = \floor s-m,\dots, \floor s$.

This is quite a non-trivial and non-obvious constraint on $f^{s,i}_m(\lambda,-\Delta_\pm)$ which will be useful as a check of our calculations. Equation \eqref{eq:FullExpressionOfThreePointFunctionNumberTwo} is one of our main results and gives us the three-point functions when removing\footnote{Recall our analysis of holographic Ward identities, where we found out that $\frac{\mu_s}{2\pi}$, not $\mu_s$, corresponds to the correct normalized source of the dual field operator.} $\frac 1{2\pi}\mu_\phi\mu_s$

\subsection{Solution of the Vasiliev Recursion Relations}
According to equation \eqref{eq:FullExpressionOfThreePointFunctionNumberTwo}, the calculation of the three-point functions is reduced to solving the Vasiliev equations \eqref{eq:VasilievEquationRecursionRelationsForScalars} recursively in order to express the auxiliary fields $C^s_{-m}$ in terms of $\phi_\pm$. This task is most easily solved by splitting it into two steps. We will first express the minimal components $C^{m+1}_{-m}$ and $C^{m+\frac 32}_{-m}$ in terms of $C^1_0$ and $C^{\frac 32}_0$, afterwards express the non-minimal components $C^{s\neq m+1,m+\frac 32}_{-m}$ in terms of $C^{m+1}_{-m}$ and $C^{m+\frac 32}_{-m}$. Combining these two solutions, we can express $C^s_{-m}$ in terms of the physical scalars $\phi_\pm$ which is what we need in equation \eqref{eq:FullExpressionOfThreePointFunctionNumberTwo}.

For the first step we need to use the $z$-equations of \eqref{eq:VasilievEquationRecursionRelationsForScalars} for the negative mode minimal components
  \begin{gather*}
    \begin{aligned}
      L^{(m+1)}_{-m,z}:&	&	&\partial C^{m+1}_{-m} + e^{\rho}\,g^{2,m+2}_3(1,-m-1)\,C^{m+2}_{-m-1} = 0,\\
      L^{(m+\frac 32)}_{-m,z}:&	&	&\partial C^{m+\frac 32}_{-m} + e^{\rho}\,g^{2,m+\frac 52}_3(1,-m-1)\,C^{m+\frac 52}_{-m-1} + e^{\rho}\,g^{2,m+2}_{\frac 52}(1,-m-1)\,C^{m+2}_{-m-1} = 0.
    \end{aligned}
  \end{gather*}
The first of these equations is readily solved
  \begin{equation}\label{eq:RecursionRelationSolution1}
    C^{m+1}_{-m} = \left(\prod_{i=1}^mg^{2,i+1}_3(1,-i)\right)^{-1}\left(-e^{-\rho}\partial\right)^mC^1_0.
  \end{equation}
The second equation is easier to solve if one considers the more general recursion relation
  \begin{equation}
    \alpha_mC^{m+\frac 32} + C^{m+\frac 52} + \beta_mC^{m+2} = 0,
  \end{equation}
which has the solution
  \begin{equation}
    C^{m+\frac 52} = \prod_{i=0}^m(-\alpha_i)C^{\frac 32} + \sum_{p=1}^{m+1}\left(\prod_{j=p}^m(-\alpha_j)\right)(-\beta_{p-1})C^{p+1}.
  \end{equation}
Putting the coefficients to
  \begin{equation}
    \alpha_m = e^{-\rho}\left(g^{2,m+\frac 52}_3(1,-m-1)\right)^{-1}\,\partial, \qquad \beta_m = \frac{g^{2,m+2}_{\frac 52}(1,-m-1)}{g^{2,m+\frac 52}_3(1,-m-1)},
  \end{equation}
and using the other solution \eqref{eq:RecursionRelationSolution1}, one can write down the solution of the second equation as
  \begin{gather}\label{eq:RecursionRelationSolution2}
    \begin{split}
      C^{m+\frac 32}_{-m} &= \left(\prod_{i=1}^mg^{2,i+\frac 32}_3(1,-i)\right)^{-1}\left(-e^{\rho}\partial\right)^mC^{\frac 32}_0 +  \sum_{p=1}^m\left(\prod_{j=p+1}^mg^{2,j+\frac 32}_3(1,-j)\right)^{-1}
	\\&\quad\times\left(\prod_{k=1}^pg^{2,k+1}_3(1,-k)\right)^{-1}\left(\frac{-g^{2,p+1}_{\frac 52}(1,-p)}{g^{2,p+\frac 32}_3(1,-p)}\right)\left(-e^{-\rho}\partial\right)^mC^1_0.
    \end{split}
  \end{gather}
One can find very similar expressions for the auxiliary fields with positive mode using the $\bar z$ equations of \eqref{eq:VasilievEquationRecursionRelationsForScalars}, these are given by

  \begin{equation}\label{eq:RecursionRelationSolution3}
    C^{m+1}_m = \left(\prod_{i=1}^mg^{i+1,2}_3(i,-1)\right)^{-1}\left(e^{-\rho}\bar{\partial}\right)^mC^1_0,
  \end{equation}
and
  \begin{gather}\label{eq:RecursionRelationSolution4}
    \begin{aligned}
      C^{m+\frac 32}_m &= \left(\prod_{i=1}^mg^{i+\frac 32,2}_3(i,-1)\right)^{-1}\left(e^{-\rho}\bar{\partial}\right)^mC^{\frac 32}_0 + \sum_{p=1}^m\left(\prod_{j=p+1}^mg^{j+\frac 32,2}_3(j,-1)\right)^{-1}\\
	&\quad\times\left(\frac{-g^{p+1,2}_{\frac 52}(p,-1)}{g^{p+\frac 32,2}_3(p,-1)}\right)\left(e^{-\rho}\bar{\partial}\right)^{m-p}C^{p+1}_p.
    \end{aligned}
  \end{gather}
Now for the second step we need to use the $\rho$-equations of \eqref{eq:VasilievEquationRecursionRelationsForScalars} given by
  \begin{equation}\label{eq:RhoEquationToSolveRecursivelyVasiliev}
    \partial_\rho C^s_m + 2C^{s-1}_m + \kappa_s\, C^{s+1}_m+ \omega_{s-1/2}\,C^{s-\frac 12}_m+\sigma_{s+\frac 12}\,C^{s+\frac 12}_m = 0,
  \end{equation}
where out of convenience we have defined the quantities
  \begin{gather}
    \begin{split}
      \kappa_s &\equiv 2g^{s+1,2}_3(m,0),\\
      \omega_{s-\frac 12} &\equiv 2 g^{s-\frac 12,2}_{\frac 32}(m,0),\\
      \sigma_{s+\frac 12} &\equiv 2 g^{s+\frac 12,2}_{\frac 52}(m,0).
    \end{split}
  \end{gather}
Note that we have suppressed the $m$ dependence since we need to solve the above equation for fixed $m$. According to the properties of the structure constants listed in appendix \ref{appendix:StructureConstantsOfHigherSpinAlgebra}, $\omega_{s-\frac 12} = 0$ for $s\in\mathbb Z+\frac 12$ and $\sigma_{s+\frac 12} = 0$ for $s\in\mathbb Z$, thus we can split \eqref{eq:RhoEquationToSolveRecursivelyVasiliev} into two slightly simpler equations\footnote{Note the exceptions $\omega_{\frac 32-\frac 12}\propto m$ and $\sigma_{1+\frac 12}\propto m$, which lead to terms of the form $m\,C^1_m$ and $m\,C^{\frac 32}_m$. Only for $m=0$ are these terms inside the wedge and thus they vanish (for $m>0$, $C^1_m = C^{\frac 32}_m =0$).}
  \begin{gather}\label{eq:RhoEquationToSolveRecursivelyVasilievTwo}
    \begin{aligned}
      \partial_\rho C^s_m + 2C^{s-1}_m + \kappa_s\,C^{s+1}_m + \omega_{s-\frac 12}\,C^{s-\frac 12}_m &= 0,\\
      \partial_\rho C^{s+\frac 12}_m+2C^{s-\frac 12}_m + \kappa_{s+\frac 12}\,C^{s+\frac 32}_m + \sigma_{s+1}\,C^{s+1}_m &=0,
    \end{aligned}\qquad s\in\mathbb Z_{\geq 1}.
  \end{gather}
Due to the $\sigma$ and $\omega$ terms these two recursion relations are coupled to each other and this makes the equations difficult to solve.\footnote{It turns out that in the case of $\sigma_s=0$, one can directly solve the recursion relations using some neat tricks. This 
solution is outlined in the first author's MSc thesis.} For our needs we can simply solve these equations recursively using computer algebra software to any desired order and then evaluate the expression \eqref{eq:FullExpressionOfThreePointFunctionNumberTwo} to explicitly find the corresponding three-point functions. Let us however make a few general and important comments. Note that the general solution will be of the form
  \begin{gather}
    \begin{aligned}\label{eq:GeneralFormOfSolutionsOfRecursionRelations}
      C^s_m &= \mathcal O_s\left(\partial_\rho\right)C^{m+1}_m + \mathcal P_s\left(\partial_\rho\right)C^{m+\frac 32}_m,\\
      C^{s+\frac 12}_m &= \tilde{\mathcal O}_s\left(\partial_\rho\right)C^{m+1}_m + \tilde{\mathcal P}_s\left(\partial_\rho\right)C^{m+\frac 32}_m,
    \end{aligned}
  \end{gather}
where the differential operators clearly do not explicitly depend on $\rho$ but only on $\partial_\rho$. In order to find the functions $\mathcal G^{s,i}_m(\lambda,\partial_\rho)$ of equation \eqref{eq:HigherSpinDeformationFunctionFDefinitnion}, we need to move the exponential factors of \eqref{eq:RecursionRelationSolution1}, \eqref{eq:RecursionRelationSolution2}, \eqref{eq:RecursionRelationSolution3}, \eqref{eq:RecursionRelationSolution4}, outside in equation \eqref{eq:GeneralFormOfSolutionsOfRecursionRelations}. Since the operators $\mathcal O_s$, $\mathcal P_s$, $\tilde{\mathcal O}_s$ and $\tilde{\mathcal P}_s$ are polynomials of $\partial_\rho$, consider the following short calculation
   \begin{gather}
     \begin{aligned}
       \partial^n_{\rho}\left(e^{-m\rho}\,\phi\right) &= \sum_{q=0}^n\begin{pmatrix}
 								    n\\q
                                                                    \end{pmatrix}
  	\partial^{n-q}_{\rho}\left(e^{-m\rho}\right)\,\partial^q_{\rho}\,\phi,\\
  						     &= \sum_{q=0}^n\begin{pmatrix}
  								     n\\q
  						                    \end{pmatrix}
  	  (-m)^{n-q}\,\partial^q_{\rho}\phi\,e^{-m\rho},\\
  						     &= \Big[\left(\partial_{\rho} - m\right)^n\phi\Big]\, e^{-m\rho},
     \end{aligned}
   \end{gather}
where we have used the binomial theorem for the differential operator in the last line. Thus if we remove by hand the exponential factors of \eqref{eq:RecursionRelationSolution1}, \eqref{eq:RecursionRelationSolution2}, \eqref{eq:RecursionRelationSolution3}, \eqref{eq:RecursionRelationSolution4}, and then shift the operators of equation \eqref{eq:GeneralFormOfSolutionsOfRecursionRelations} by
  \[	\partial_\rho\rightarrow\partial_\rho - m,	\]
we will find the functions $\mathcal G^{s,i}_m(\lambda,\partial_\rho)$. This is an important detail to remember when implementing these recursion relations \eqref{eq:RhoEquationToSolveRecursivelyVasilievTwo} in a computer algebra package.

As a final remark, let us note that the $\rho$ and $\partial$ dependence of the auxiliary fields are of the form\footnote{Recall that we always set one of the scalars $\phi_\pm$ to zero.} $C^s_{-m}\sim e^{-|m|\rho}\,A\big(\lambda,\partial_{\rho}\big)\,\partial^m\phi_i$ as claimed and used earlier.

\subsection{Results for Three-Point Functions from the bulk}\label{Subsection:BulkResults}

We can finally calculate the three-point functions by using equation \eqref{eq:FullExpressionOfThreePointFunctionNumberTwo}, removing the $\frac 1{2\pi}\mu_\phi\mu_s$ factor, together with the solution of the above recursion relations. It is however difficult to proceed analytically partly because we do not have a general closed formula for the recursion relations, but mainly because the structure constants of $\mathcal{SB}[\mu]$ are complicated expressions and it is hard to rewrite the whole thing as simple functions of $\lambda$. We will therefore proceed by explicitly calculating the different three-point functions for low spin $s$, then extrapolating the result to arbitrary $s$. These closed expressions can then be checked on a computer for a large number of spins $s$.

Let us briefly comment on some consistency checks. We have checked that the constraints \eqref{eq:ConstraintsOnThefFunction} are satisfied for a wide range of values $s$. Remarkably, if we modify the expression \eqref{eq:HigherSpinDeformationFunctionFDefinitnion}, even slightly, then the constraint \eqref{eq:ConstraintsOnThefFunction} will no longer be satisfied. Furthermore it turns out that the expression \eqref{eq:FullExpressionOfThreePointFunctionNumberTwo} for the three-point functions exactly ends up having the correct $(z_1,z_2,z_3)$-dependence which is required by conformal symmetry \eqref{eq:ThreePointFunctionzDependenceRequiredByConformalSymmetry}, but is not manifest from \eqref{eq:FullExpressionOfThreePointFunctionNumberTwo} at all. Here we also observe that even the smallest changes of the equations \eqref{eq:HigherSpinDeformationFunctionFDefinitnion} or \eqref{eq:FullExpressionOfThreePointFunctionNumberTwo} will result in ``three-point functions'' with complicated $(z_1,z_2,z_3)$-dependence and the result will not respect conformal symmetry \eqref{eq:ThreePointFunctionzDependenceRequiredByConformalSymmetry}. 
The fact that these and other similar, highly non-trivial, checks work out is quite remarkable and provides confidence in 
the consistency and robustness of our results.

Since all three-point functions we are considering are of the form
  \[
	\average[\big]{\mathcal O_\Delta(z_1,\bar z_1)\bar{\mathcal O}_\Delta(z_2,\bar z_2)J^{(s)}(z_3)} = \average[\big]{\mathcal O_\Delta\bar{\mathcal O}_\Delta J^{(s)}}\left(\frac{z_{12}}{z_{13}z_{23}}\right)^s\,\average[\big]{\mathcal O_\Delta(z_1,\bar z_1)\bar{\mathcal O}_\Delta(z_2,\bar z_2)},
  \]
we will use the notation $\average[\big]{\mathcal O_\Delta\bar{\mathcal O}_\Delta J^{(s)}}$ to denote the coefficients. Let us take the dual operator of $\phi_+$ with conformal weight $\Delta_+ = 2(1-\lambda)$. By solving the recursion relations above and following the detailed procedure developed in this chapter, equation \eqref{eq:FullExpressionOfThreePointFunctionNumberTwo} gives us the following coefficients for low spin
  \begin{gather}
    \begin{aligned}
      \average[\big]{\mathcal O_{\Delta_+}^{\mathcal B}\bar{\mathcal O}_{\Delta_+}^{\mathcal B} W^{2+}} &= -(\lambda-1),\\
      \average[\big]{\mathcal O_{\Delta_+}^{\mathcal B}\bar{\mathcal O}_{\Delta_+}^{\mathcal B} W^{3+}} &= -\frac 13\, (\lambda-1)\,(2\lambda-3),\\
      \average[\big]{\mathcal O_{\Delta_+}^{\mathcal B}\bar{\mathcal O}_{\Delta_+}^{\mathcal B} W^{4+}} &= -\frac 15\, (\lambda -2)\,(\lambda-1)\,(2\lambda-3),\\
      \average[\big]{\mathcal O_{\Delta_+}^{\mathcal B}\bar{\mathcal O}_{\Delta_+}^{\mathcal B} W^{5+}} &= -\frac 2{35}\, (\lambda -2)\,(\lambda-1)\,(2\lambda-5)\,(2\lambda-3),\\
      \average[\big]{\mathcal O_{\Delta_+}^{\mathcal B}\bar{\mathcal O}_{\Delta_+}^{\mathcal B} W^{6+}} &= -\frac 2{63}\, (\lambda -3)\,(\lambda -2)\,(\lambda-1)\,(2\lambda-5)\,(2\lambda-3).
    \end{aligned}
  \end{gather}
Note that $W^{2+}(z)$ is the holomorphic part of the energy-momentum tensor and therefore the coefficient of the three-point function must be the holomorphic conformal weight of $\mathcal O^{\mathcal B}_{\Delta_+}$ which is $h_+ = 1-\lambda$, see equation \eqref{eq:ReviewSectionCFTPrimariesFirstMultiplet} and \eqref{eq:CosetChiralFieldConformalWeights}. Encouragingly this is exactly what we find. Let us also show a few low-spin results with the same scalar but the other bosonic higher-spin current $W^{s-}$
  \begin{gather}
    \begin{aligned}
      \average[\big]{\mathcal O_{\Delta_+}^{\mathcal B}\bar{\mathcal O}_{\Delta_+}^{\mathcal B} W^{2-}} &= -\frac 13\,(\lambda-1)\,(2\lambda+1),\\
      \average[\big]{\mathcal O_{\Delta_+}^{\mathcal B}\bar{\mathcal O}_{\Delta_+}^{\mathcal B} W^{3-}} &= -\frac 2{15}\,(\lambda-1)\,(\lambda+1)\,(2\lambda -3),\\
      \average[\big]{\mathcal O_{\Delta_+}^{\mathcal B}\bar{\mathcal O}_{\Delta_+}^{\mathcal B} W^{4-}} &= -\frac 1{35}\,(\lambda-2)\,(\lambda-1)\,(2\lambda -3)\,(2\lambda+3),\\
      \average[\big]{\mathcal O_{\Delta_+}^{\mathcal B}\bar{\mathcal O}_{\Delta_+}^{\mathcal B} W^{5-}} &= -\frac 4{315}\,(\lambda-2)\,(\lambda-1)\,(\lambda +2)\,(2\lambda-5)\,(2\lambda -3),\\
      \average[\big]{\mathcal O_{\Delta_+}^{\mathcal B}\bar{\mathcal O}_{\Delta_+}^{\mathcal B} W^{6-}} &= -\frac 2{693}\,(\lambda-3)\,(\lambda-2)\,(\lambda -1)\,(2\lambda-5)\,(2\lambda -3)\,(2\lambda+5).
    \end{aligned}
  \end{gather}
Amazingly it turns out that all three-point functions factorize as the above examples and thus make it easy for us to guess the correct closed form expression for all spin. For the CFT dual fields corresponding to $\tilde \phi_\pm$, we need to multiply by a factor of $(-1)^s$ due to the different coupling to the higher-spin fields \eqref{eq:VasilievEquationMatterCoupledToGaugeField}. The general expressions are given by
  \begin{gather}\label{eq:ThreePointFunctionResultsFromBulkPlus}
    \begin{split}
	\average[\big]{\mathcal O_{\Delta_+}^{\mathcal B}\bar{\mathcal O}_{\Delta_+}^{\mathcal B} W^{s+}} &= (-1)^s\frac{\Gamma^2(s)}{\Gamma(2s-1)}\,\frac{\Gamma(s-2\lambda+1)}{\Gamma(2-2\lambda)},\\
	\average[\big]{\mathcal O_{\Delta_-}^{\mathcal B}\bar{\mathcal O}_{\Delta_-}^{\mathcal B} W^{s+}} &= (-1)^s\frac{\Gamma^2(s)}{\Gamma(2s-1)}\,\frac{\Gamma(s-2\lambda)}{\Gamma(1-2\lambda)},\\
	\average[\big]{\tilde{\mathcal O}_{\Delta_+}^{\mathcal B}\bar{\tilde{\mathcal O}}_{\Delta_+}^{\mathcal B} W^{s+}} &= (-1)^{s-1}\frac{\Gamma^2(s)}{\Gamma(2s-1)}\,\frac{\Gamma(-2\lambda+1)}{\Gamma(-2\lambda-s+2)},\\
	\average[\big]{\tilde{\mathcal O}_{\Delta_-}^{\mathcal B}\bar{\tilde{\mathcal O}}_{\Delta_-}^{\mathcal B} W^{s+}} &= (-1)^{s-1}\frac{\Gamma^2(s)}{\Gamma(2s-1)}\,\frac{\Gamma(-2\lambda)}{\Gamma(-2\lambda-s+1)},
    \end{split}
  \end{gather}
We have checked these closed-form expressions with our actual calculation for many spins and find perfect match. It is 
possible to combine these results into more unified formulas which depend  only on $s$, the holomorphic conformal weights and the
type of the fields involved, as 
\be
\begin{split}
\average[\big]{\mathcal O_{h}^{\mathcal B}\bar{\mathcal O}_{h}^{\mathcal B} W^{s+}} &= (-1)^s\frac{\Gamma^2(s)}{\Gamma(2s-1)}\,
\frac{\Gamma(s+2h-1)}{\Gamma(2h)}\;,\\
\average[\big]{\tilde{\mathcal O}_{h}^{\mathcal B}\bar{\tilde{\mathcal O}}_{h}^{\mathcal B} W^{s+}} &= 
\frac{\Gamma^2(s)}{\Gamma(2s-1)}\,\frac{\Gamma(s+2h-1)}{\Gamma(2h)}\;.
\end{split}
\ee
Comparing these general formulas with the non-supersymmetric results of \cite{Ammon:2011ua}, 
and accounting for the different conformal weights in that case ($h_\pm=\half(1\pm \lambda)$), we find perfect agreement (up to
a normalization-dependent factor of $-1/(2\pi)$). Note furthermore that, as anticipated at the end of section 
\ref{eq:ModifiedVasilievFormalism}, we can obtain the above results directly from those of \cite{Ammon:2011ua} by 
substituting $\lambda_{AKP}=1-2\lambda$ (for normal quantization) for $O_{\Delta_+}^{\mathcal B}$  
and $\lambda_{AKP}=2\lambda$ (for alternate quantization) for $O_{\Delta_-}^{\mathcal B}$. 

We can follow the same procedure to find the three-point functions containing the other bosonic higher-spin fields,
which are not present in the non-supersymmetric case:
  \begin{gather}\label{eq:ThreePointFunctionResultsFromBulkMinus}
    \begin{split}
	\average[\big]{\mathcal O_{\Delta_+}^{\mathcal B}\bar{\mathcal O}_{\Delta_+}^{\mathcal B} W^{s-}} &= (-1)^{s-1}\frac{\Gamma^2(s)}{\Gamma(2s-1)}\,\frac{\Gamma(s-2\lambda+1)}{\Gamma(2-2\lambda)}\,\frac{s-1+2\lambda}{2s-1},\\
	\average[\big]{\mathcal O_{\Delta_-}^{\mathcal B}\bar{\mathcal O}_{\Delta_-}^{\mathcal B} W^{s-}} &= (-1)^s\frac{\Gamma^2(s)}{\Gamma(2s-1)}\,\frac{\Gamma(s-2\lambda)}{\Gamma(1-2\lambda)}\,\frac{s-2\lambda}{2s-1},\\
	\average[\big]{\tilde{\mathcal O}_{\Delta_+}^{\mathcal B}\bar{\tilde{\mathcal O}}_{\Delta_+}^{\mathcal B} W^{s-}} &=(-1)^s\frac{\Gamma^2(s)}{\Gamma(2s-1)}\,\frac{\Gamma(-2\lambda+1)}{\Gamma(-2\lambda-s+2)}\,\frac{s-1+2\lambda}{2s-1},\\
	\average[\big]{\tilde{\mathcal O}_{\Delta_-}^{\mathcal B}\bar{\tilde{\mathcal O}}_{\Delta_-}^{\mathcal B} W^{s-}} &= (-1)^{s-1}\frac{\Gamma^2(s)}{\Gamma(2s-1)}\,\frac{\Gamma(-2\lambda)}{\Gamma(-2\lambda-s+1)}\,\frac{s-2\lambda}{2s-1}.
    \end{split}
  \end{gather}
The coefficients \eqref{eq:ThreePointFunctionResultsFromBulkPlus} and \eqref{eq:ThreePointFunctionResultsFromBulkMinus} are our main results from the bulk calculation. 

 We notice that the coefficients of the same primaries with the $W^{s+}$ and $W^{s-}$ currents are very closely 
related. Although the reason for this similarity is not very clear from the bulk calculation, it is obvious 
from the boundary theory perspective, as we will see in the following section. Note that the results for the $W^{s-}$ 
case cannot be extracted directly from those of \cite{Ammon:2011ua}, but this is an artifact of our basis. In the
basis adapted to the projectors $\Pi_\pm$, the results would be expected to be directly comparable.

It is straightforward to generalize the above in order to obtain correlation functions containing fermions. 
For fermionic matter, one would need to set the scalar fields to zero in (\ref{eq:ExpansionOfGaugeAndMatterFieldInTermsOfSuperHigherSpinAlgebra}) while
keeping the fermionic ones. In order to include a fermionic higher-spin current, one would need to keep
only a particular fermionic higher-spin generator in (\ref{eq:GeneralizedAnsatzForGaugeFieldSinceWeNeedBoundarySourceTerms}).
Then the procedure in this section can be repeated with minor modifications. This is currently under investigation.

\section{Three-point functions from the CFT} \label{Boundary}
We will now switch gears and consider the same problem from the boundary CFT point of view. Recall that the dual CFT is defined as a double scaling limit $N,k\rightarrow\infty$ of a WZW coset \eqref{eq:KazamaSuzukiCoset}, so a direct calculation would require us to first calculate the three-point functions for finite $N$ and $k$ and then take the 't Hooft limit, so that it can be compared to the bulk calculation. It would clearly be much simpler to directly find the results in the 't Hooft limit.

All of our three-point functions take the form \eqref{eq:ThreePointFunctionzDependenceRequiredByConformalSymmetry}, where the coefficients are given by the leading order pole of the OPE
  \begin{equation}\label{eq:LeadingOrderSingularityAndThreePointFunctions}
    J^{(s)}(z)\mathcal O_\Delta(w,\bar w) \sim \frac{A(s)}{(z-w)^s}\,\mathcal O_\Delta(w,\bar w) + \dots.
  \end{equation}
If we use a standard Laurent expansion $J^{(s)}(z) = \sum_n J^{(s)}_n\,z^{-n-s}$, we can turn this into
  \begin{equation}\label{eq:HigherSpinZeroModeOfPrimaryFields}
    J^{(s)}_0\ket{\mathcal O_\Delta} = A(s)\ket{\mathcal O_\Delta}.
  \end{equation}
Recently \cite{Candu:2012tr} has put forward strong arguments in favor of the claim that in the 't Hooft limit the symmetry algebra of the Kazama-Suzuki model extends to $\mathcal{SW}_\infty[\lambda]$. 
Thus the three-point functions can be found by calculating the higher-spin zero modes of $\mathcal O_\Delta$, which is a problem in representation theory of $\mathcal{SW}_{\infty}[\lambda]$. This is in general not such a simple problem due to non-linearities of the algebra, especially for arbitrary central charge $c$. In the 't Hooft limit however, it turns out that all non-linear terms vanish for supercommutators between elements in the wedge. In other words in the $c\rightarrow\infty$ limit, $\shs[\lambda]$ becomes a subalgebra of $\mathcal{SW}_\infty[\lambda]$ and by the arguments of \cite{Gaberdiel:2011wb} these zero modes can be calculated purely from representation theory of $\shs[\lambda]$.\footnote{See \cite{Chang:2011mz,Ammon:2011ua} for the same approach in the non-supersymmetric case.}

Instead of analyzing the representation theory of $\shs[\lambda]$, it is much simpler to generate the relevant representations field theoretically by constructing a CFT with $\shs[\lambda]$ as a subalgebra. The simplest CFT's one can imagine are free CFT's, in which there are an infinite number of higher-spin conserved currents.

Following \cite{Bergshoeff:1991dz}, we will use the simple ghost system
\be
S=\frac{1}{\pi}\int\diff^2 z\left\{ b\pbar c +\beta \pbar \gamma + \bt \p \ct + \betat \p \gammat\right\}
\ee
which has the free field OPE's:
\be\label{eq:FreeFieldOPEs}
\gamma(z)\beta(w) \sim \frac{1}{z-w} \;\;,\quad {\text{and}}\quad c(z)b(w)\sim\frac{1}{z-w}
\ee
and similarly for the tilded fields. Here $b$, $c$, $\tilde b$ and $\tilde c$ are fermionic while $\beta$, $\gamma$, $\tilde{\beta}$ and $\tilde{\gamma}$ are bosonic. It was shown in \cite{Bergshoeff:1991dz} that this free CFT has an infinite number of conserved currents which together form the $\mathcal N=2$ linear $sw_\infty[\lambda]\oplus sw_\infty[\lambda]$ algebra. Although this is of course not equivalent to the $\mathbb CP^N$ Kazama-Suzuki model and these theories do not even have the non-linear $\mathcal{SW}_\infty[\lambda]\oplus\mathcal{SW}_\infty[\lambda]$ algebra in common, they both have an $\shs[\lambda]\oplus\shs[\lambda]$ closed subalgebra. This implies that if we can construct primary fields with the correct conformal weights in this free theory, then the coefficients of the leading order pole \eqref{eq:LeadingOrderSingularityAndThreePointFunctions} would exactly correspond to the higher-spin zero mode and thereby the coefficients of three-point functions of the Kazama-Suzuki CFT in the 't Hooft limit.

The conformal weights of the fields are given by
\begin{center}
\begin{tabular}{c|cccccccc}%\hline
& $b$ &$c$ & $\beta$ &$\gamma$ &$\bt$ &$\ct$ &$\betat$& $\gammat$\\ \hline
$h$ & $\lambda+\mhalf$ & $\mhalf-\lambda $&$ \lambda$ & $1-\lambda$& $0$& $0$& $0$& $0$\\ 
$\hb$ & $0$&$0$&$0$&$0$&$\lambda+\mhalf$&$\mhalf-\lambda$ & $\lambda$&$1-\lambda$%\\ \hline
\end{tabular}
\end{center}
Remarkably, this is exactly the same as the coset primaries discussed in section \ref{Review}.

We will use these fields to construct CFT operators that are dual to the bulk fields
$\phi_\pm,\phit_\pm,\psi_\pm,\psit_\pm$.
Recall that \cite{Prokushkin:1998bq,Prokushkin:1998vn,Creutzig:2011fe} the bulk fields are 
arranged in multiplets of $\Ncal=2$ supersymmetry: 
\be
\big(\phi_+,\psi_{\pm},\phi_-\big) \qquad \text{and}\qquad \big(\phit_+,\psit_\pm,\phit_-\big),
\ee
where the scalars appearing in each multiplet have different masses,
$(M^B_+)^2=(\tilde{M}^B_+)^2=-4\lambda(1-\lambda)$ and $(M^B_-)^2=(\tilde{M}^B_-)^2=-1+4\lambda^2$, and are also oppositely 
quantized ($\phi_+$ and $\phit_-$ have the usual quantization, $\phi_-$ and $\phit_+$ the 
alternative one). 

Identifying these fields with the coset fields, we can construct the dual fields as discussed in section \ref{Review}
  \begin{gather}\label{eq:CFTPrimariesFirstMultiplet}
    \begin{aligned}
      \mathcal O^{\mathcal B}_{\Delta_+}(z,\zbar)&=\gamma(z)\otimes \tilde\gamma(\zbar),&\quad \mathcal O^{\mathcal F}_{\Delta_+}(z,\zbar)&=c(z)\otimes \tilde\gamma(\zbar), \\
      \mathcal O^{\mathcal B}_{\Delta_-}(z,\zbar)&=c(z)\otimes \tilde c(\zbar),& \quad \mathcal O^{\mathcal F}_{\Delta_-}(z,\zbar)&=\gamma(z) \otimes \tilde c(\zbar),
    \end{aligned}
  \end{gather}
and
  \begin{gather}\label{eq:CFTPrimariesSecondMultiplet}
    \begin{aligned}
      \tilde{\mathcal O}^{\mathcal B}_{\Delta_+}(z,\zbar)&=\beta(z)\otimes \tilde\beta(\zbar),&\quad \tilde{\mathcal O}^{\mathcal F}_{\Delta_+}(z,\zbar)&=b(z) \otimes \tilde\beta(\zbar), \\
      \tilde{\mathcal O}^{\mathcal B}_{\Delta_-}(z,\zbar)&=b(z)\otimes \tilde b(\zbar), &\quad \tilde{\mathcal O}^{\mathcal F}_{\Delta_-}(z,\zbar)&=\beta(z)\otimes \tilde b(\zbar). 
    \end{aligned}
  \end{gather}
The scaling dimensions of these fields $\Delta=h+\hb$
precisely match the dimensions corresponding to the bulk fields with the appropriate quantization, as discussed earlier.

The higher-spin currents corresponding to the linear $sw_\infty[\lambda]\oplus sw_\infty[\lambda]$ algebra are given by \cite{Bergshoeff:1991dz}:
\be\label{eq:BosonicHigherSpinCurrentPlus}
V_\lambda^{(s)+}(z)=\sum_{i=0}^{s-1} a^i(s,\lambda) \p^{s-1-i}\left\{(\p^i\beta)\gamma\right\}
+\sum_{i=0}^{s-1} a^i(s,\lambda+\mhalf) \p^{s-1-i}\left\{(\p^ib)c\right\},
\ee
\be\label{eq:BosonicHigherSpinCurrentMinus}
V_\lambda^{(s)-}(z)=-\frac{s-1+2\lambda}{2s-1}\sum_{i=0}^{s-1} a^i(s,\lambda)\p^{s-1-i}\left\{(\p^i\beta)\gamma\right\}
+\frac{s-2\lambda}{2s-1}\sum_{i=0}^{s-1} a^i(s,\lambda+\mhalf)\p^{s-1-i}\left\{(\p^i b) c\right\},
\ee
and
\be\label{eq:FermionicHigherSpinCurrents}
Q_\lambda^{(s)\pm}(z)=\sum_{i=0}^{s-1} \alpha^i(s,\lambda) \p^{s-1-i}\left\{(\p^i\beta)c\right\}
\mp \sum_{i=0}^{s-2} \beta^i(s,\lambda) \p^{s-2-i}\left\{(\p^i b) \gamma \right\},
\ee
and similarly for the anti-holomorphic sector. The coefficients are given by
  \begin{gather}\label{eq:DefinitionOfTheaalphabetaFunctions}
    \begin{aligned}
      a^i(s,\lambda) &= \begin{pmatrix}
			  s-1\\i
		      \end{pmatrix}
	\frac{(-2\lambda-s+2)_{s-1-i}}{(s+i)_{s-1-i}}, \quad & \quad &0\leq i \leq s-1,\\
      \alpha^i(s,\lambda) &= \begin{pmatrix}
			  s-1\\i
		      \end{pmatrix}
	\frac{(-2\lambda-s+2)_{s-1-i}}{(s+i-1)_{s-1-i}}, \quad & \quad &0\leq i \leq s-1,\\
      \beta^i(s,\lambda) &= \begin{pmatrix}
			  s-2\\i
		      \end{pmatrix}
	\frac{(-2\lambda-s+2)_{s-2-i}}{(s+i)_{s-2-i}}, \quad & \quad &0\leq i \leq s-2.
    \end{aligned}
  \end{gather}
These currents are normalized such that their Laurent modes (when restricting to the wedge) correspond to the $\shs[\lambda]$ generators \eqref{eq:ListOfHigerSpinAlgebraGenerators} in the exact same basis \cite{Bergshoeff:1991dz}. Thus the higher-spin zero modes of the dual fields \eqref{eq:HigherSpinZeroModeOfPrimaryFields}, and thereby three-point functions should be directly comparable.

\subsection{Operator product expansions}

In order to compute three-point functions involving the higher-spin currents we need to compute the coefficient of the leading order pole of the OPE between higher-spin currents and the primaries \eqref{eq:CFTPrimariesFirstMultiplet} and \eqref{eq:CFTPrimariesSecondMultiplet}. It is straightforward to do this using \eqref{eq:FreeFieldOPEs} and the form of the higher-spin currents given in \eqref{eq:BosonicHigherSpinCurrentPlus}, \eqref{eq:BosonicHigherSpinCurrentMinus} and \eqref{eq:FermionicHigherSpinCurrents}, we will list the result here. For $V^{(s)+}_\lambda$ we have
  \begin{gather}
    \begin{split}
      V_\lambda^{(s)+}(z) \beta(w) &\sim a^0(s,\lambda)\frac{(-1)^{s-1}(s-1)!}{(z-w)^s}\beta(w)+\cdots\;,\\
      V_\lambda^{(s)+}(z) b(w) &\sim a^0(s,\lambda+\mhalf)\frac{(-1)^{s-1}(s-1)!}{(z-w)^s} b(w)+\cdots\;,\\
      V_\lambda^{(s)+}(z) \gamma(w) &\sim \left(\sum_{i=0}^{s-1}a^i(s,\lambda)\right)~\frac{(-1)^s(s-1)!}{(z-w)^s}\gamma(w)+\cdots\;,\\
      V_\lambda^{(s)+}(z) c(w)&\sim\left(\sum_{i=0}^{s-1} a^i(s,\lambda+\mhalf)\right)\frac{(-1)^s (s-1)!}{(z-w)^s} c(w)+\cdots\;.\\
    \end{split}
  \end{gather}
In a similar manner we find that the OPE's involving the $V_\lambda^{(s)-}$ currents are given by
  \begin{gather}
    \begin{split}
      V_\lambda^{(s)-}(z) \beta(w)&\sim \frac{s-1+2\lambda}{2s-1} a^0(s,\lambda) \frac{(-1)^s (s-1)!}{(z-w)^s} \beta(w)+\cdots\;,\\
      V_\lambda^{(s)-}(z) b(w)&\sim \frac{s-2\lambda}{2s-1} a^0(s,\lambda+\mhalf)\frac{(-1)^{s-1}(s-1)!}{(z-w)^s} b(w) + \cdots\;,\\
      V_\lambda^{(s)-}(z) \gamma(w)&\sim \frac{s-1+2\lambda}{2s-1}\left(\sum_{i=0}^{s-1} a^i(s,\lambda)\right) \frac{(-1)^{s-1} (s-1)!}{(z-w)^s} \gamma(w)+\cdots\;,\\
      V_\lambda^{(s)-}(z) c(w)&\sim \frac{s-2\lambda}{2s-1}\left(\sum_{i=0}^{s-1} a^i(s,\lambda+\mhalf)\right) \frac{(-1)^s (s-1)!}{(z-w)^s} c(w)+\cdots\;.
    \end{split}
  \end{gather}
Finally for the fermionic higher-spin currents $Q^{(s)\pm}_\lambda$ we find
  \begin{gather}
    \begin{split}
      Q^{(s)\pm}_\lambda(z)\beta(w) &\sim\mp\beta^0(s,\lambda)\frac{(-1)^s (s-2)!}{(z-w)^{s-1}}b(w) + \cdots\;,\\
      Q^{(s)\pm}_\lambda(z)b(w) &\sim\alpha^0(s,\lambda)\frac{(-1)^{s-1} (s-1)!}{(z-w)^s}\beta(w) + \cdots\;,\\
      Q^{(s)\pm}_\lambda(z)\gamma(w) &\sim\left(\sum_{i=0}^{s-1}\alpha^s(s,\lambda)\right)\frac{(-1)^s (s-1)!}{(z-w)^s}c(w) + \cdots\;,\\
      Q^{(s)\pm}_\lambda(z)c(w) &\sim\mp\left(\sum_{i=0}^{s-2}\beta^i(s,\lambda)\right)\frac{(-1)^s (s-2)!}{(z-w)^{s-1}}\gamma(w) + \cdots\;.
    \end{split}
  \end{gather}
In order to be able to compare the CFT three-point functions with the bulk results \eqref{eq:ThreePointFunctionResultsFromBulkPlus} and \eqref{eq:ThreePointFunctionResultsFromBulkMinus}, we will write the coefficients in the following form
  \begin{gather}
    \begin{split}
    a^0(s,\lambda)(s-1)! &= \frac{\Gamma(s)^2}{\Gamma(2s-1)}\frac{\Gamma(-2\lambda+1)}{\Gamma(-2\lambda-s+2)},\\
    \beta^0(s,\lambda)(s-2)!&= \frac{\Gamma(s-1)\Gamma(s)}{\Gamma(2s-2)}\frac{\Gamma(-2\lambda)}{\Gamma(-2\lambda-s+2)},\\
    \alpha^0(s,\lambda)(s-1)! &= \frac{\Gamma(s)\Gamma(s-1)}{\Gamma(2s-2)}\frac{\Gamma(-2\lambda+1)}{\Gamma(-2\lambda-s+2)}.
    \end{split}
  \end{gather}
Furthermore it is straightforward to perform the necessary sums over the coefficients, which results in
  \begin{gather}
    \begin{split}
      \sum_{i=0}^{s-1} a^i(s,\lambda)&=\frac{4^{1-s}\sqrt{\pi}\,\Gamma(1+s-2\lambda)}{\Gamma(s-\mhalf)\Gamma(2-2\lambda)}
=\frac{\Gamma(s)}{\Gamma(2s-1)}\frac{\Gamma(1+s-2\lambda)}{\Gamma(2-2\lambda)},\\
      \sum_{i=0}^{s-2} \beta^i(s,\lambda) &= \frac{2^{3-2s} \sqrt{\pi}\,(s-1)\,\Gamma(s-2\lambda)}{\Gamma(s-\mhalf)\Gamma(2-2\lambda)} = 2\frac{\Gamma(s)(s-1)}{\Gamma(2s-1)}\frac{\Gamma(s-2\lambda)}{\Gamma(2-2\lambda)},\\
      \sum_{i=0}^{s-1}\alpha^i(s,\lambda) &= \frac{(-1)^{s-1} 2^{3-2s} \sqrt{\pi}\,\Gamma(2\lambda)}{\Gamma(s-\mhalf)\Gamma(1-s+2\lambda)} = (-1)^{s-1} 2\frac{\Gamma(s)}{\Gamma(2s-1)}\frac{\Gamma(2\lambda)}{\Gamma(1-s+2\lambda)},
    \end{split}
  \end{gather}
for $s>1$.

\subsection{Bosonic three-point correlators from the CFT} \label{BoundarySection}
We now have all the necessary ingredients to compute all three-point correlators of two 
bosonic or fermionic operators with a spin-$s$ current. In this section we list all the bosonic three-point functions, using the notation used in section \ref{Subsection:BulkResults}.

\paragraph{First Multiplet  with $V_\lambda^{(s)+}$}\mbox{}
  \begin{gather}
    \begin{split}
	\average[\big]{\mathcal O_{\Delta_+}^{\mathcal B}\bar{\mathcal O}_{\Delta_+}^{\mathcal B} V^{(s)+}_\lambda} &= (-1)^s\frac{\Gamma^2(s)}{\Gamma(2s-1)}\,\frac{\Gamma(s-2\lambda+1)}{\Gamma(2-2\lambda)},\\
	\average[\big]{\mathcal O_{\Delta_-}^{\mathcal B}\bar{\mathcal O}_{\Delta_-}^{\mathcal B} V^{(s)+}_\lambda} &= (-1)^s\frac{\Gamma^2(s)}{\Gamma(2s-1)}\,\frac{\Gamma(s-2\lambda)}{\Gamma(1-2\lambda)}.
    \end{split}
  \end{gather}
\paragraph{Second multiplet  with $V_\lambda^{(s)+}$}\mbox{}
  \begin{gather}
    \begin{split}
	\average[\big]{\tilde{\mathcal O}_{\Delta_+}^{\mathcal B}\bar{\tilde{\mathcal O}}_{\Delta_+}^{\mathcal B} V^{(s)+}_\lambda} &= (-1)^{s-1}\frac{\Gamma^2(s)}{\Gamma(2s-1)}\,\frac{\Gamma(-2\lambda+1)}{\Gamma(-2\lambda-s+2)},\\
	\average[\big]{\tilde{\mathcal O}_{\Delta_-}^{\mathcal B}\bar{\tilde{\mathcal O}}_{\Delta_-}^{\mathcal B} V^{(s)+}_\lambda} &= (-1)^{s-1}\frac{\Gamma^2(s)}{\Gamma(2s-1)}\,\frac{\Gamma(-2\lambda)}{\Gamma(-2\lambda-s+1)}.
    \end{split}
  \end{gather}
\paragraph{First multiplet with $V_\lambda^{(s)-}$}\mbox{}
  \begin{gather}
   \begin{split}
	\average[\big]{\mathcal O_{\Delta_+}^{\mathcal B}\bar{\mathcal O}_{\Delta_+}^{\mathcal B} V^{(s)-}_\lambda} &= (-1)^{s-1}\frac{\Gamma^2(s)}{\Gamma(2s-1)}\,\frac{\Gamma(s-2\lambda+1)}{\Gamma(2-2\lambda)}\,\frac{s-1+2\lambda}{2s-1},\\
	\average[\big]{\mathcal O_{\Delta_-}^{\mathcal B}\bar{\mathcal O}_{\Delta_-}^{\mathcal B} V^{(s)-}_\lambda} &= (-1)^s\frac{\Gamma^2(s)}{\Gamma(2s-1)}\,\frac{\Gamma(s-2\lambda)}{\Gamma(1-2\lambda)}\,\frac{s-2\lambda}{2s-1}.
    \end{split}
  \end{gather}
\paragraph{Second multiplet with $V_\lambda^{(s)-}$}\mbox{}
  \begin{gather}
    \begin{split}
	\average[\big]{\tilde{\mathcal O}_{\Delta_+}^{\mathcal B}\bar{\tilde{\mathcal O}}_{\Delta_+}^{\mathcal B} V^{(s)-}_\lambda} &=(-1)^s\frac{\Gamma^2(s)}{\Gamma(2s-1)}\,\frac{\Gamma(-2\lambda+1)}{\Gamma(-2\lambda-s+2)}\,\frac{s-1+2\lambda}{2s-1},\\
	\average[\big]{\tilde{\mathcal O}_{\Delta_-}^{\mathcal B}\bar{\tilde{\mathcal O}}_{\Delta_-}^{\mathcal B} V^{(s)-}_\lambda} &= (-1)^{s-1}\frac{\Gamma^2(s)}{\Gamma(2s-1)}\,\frac{\Gamma(-2\lambda)}{\Gamma(-2\lambda-s+1)}\,\frac{s-2\lambda}{2s-1}.
    \end{split}
  \end{gather}
Comparing with the bulk computation of the same quantities we find precise agreement. This provides a non-trivial
check of the $\Ncal=2$ proposal of \cite{Creutzig:2011fe}.

\subsection{Fermionic three-point correlators from the CFT}

The above methods can also be used to compute boundary three-point functions involving
fermions. It is immediately clear that the coefficients of correlators involving
two fermionic operators and one holomorphic bosonic higher-spin current will be the
same as those of the bosonic correlators of operators that share the same chiral 
part. For instance, we can see that
\be
\average[\big]{\mathcal O_{\Delta_+}^{\mathcal F}\bar{\mathcal O}_{\Delta_+}^{\mathcal F} V^{(s)\pm}_\lambda} 
=\average[\big]{\mathcal O_{\Delta_-}^{\mathcal B}\bar{\mathcal O}_{\Delta_-}^{\mathcal B} V^{(s)\pm}_\lambda}.
\ee
On the other hand, the coefficients of the three-point functions involving one bosonic primary, 
one fermionic primary and a fermionic higher-spin current will be different. As an example, we find 
\be
\average[\big]{\mathcal O_{\Delta_+}^{\mathcal F}\bar{\mathcal O}_{\Delta_+}^{\mathcal B} Q^{(s)\pm}_\lambda}
=\pm 2 (-1)^s\frac{\Gamma(s)^2}{\Gamma(2s-1)}\frac{\Gamma(s-2\lambda)}{\Gamma(2-2\lambda)}.
\ee
It would clearly be interesting to compute the above fermionic coefficients from the bulk side of the
duality. This computation will require a straighforward  generalization of the discussion in section \ref{Bulk} and 
we hope to report on it in the near future.

\section{Conclusions}

 In this work we considered the proposal of \cite{Creutzig:2011fe} that the $\Ncal=2$ 
Prokushkin-Vasiliev theory on $\AdS_3$ is dual to a $\mathbb CP^N$ Kazama-Suzuki model 
with the non-linear chiral algebra $\mathcal{SW}_\infty[\lambda]$. In the 't Hooft limit,
we showed exact matching between three-point functions involving two bulk scalars and 
one bosonic higher-spin field as computed from the bulk and the same quantities computed
in the dual CFT. Since the correlation functions in this class only depend on the linear
$\shs[\lambda]$ algebra, they can be computed in any CFT that shares this symmetry,
and we chose to compute them in a free-field ghost CFT. This greatly simplified the boundary
side of the computation. 

 Along the way, we also performed a systematic analysis of the holographic OPE's of the
conserved higher-spin currents from the bulk theory. This demonstrates how the $\Ncal=2$
$\mathcal SW_{\infty}[\lambda]$ symmetry arises as an asymptotic symmetry and provides an 
alternative (holographic) derivation to that in \cite{Hanaki:2012yf}. In particular, it allows us 
to precisely fix the normalizations of the source terms in the boundary CFT.

 In \cite{Creutzig:2011fe}, a specific gluing of coset chiral states was proposed as dual
to the bulk fields (see \eqref{eq:ReviewSectionCFTPrimariesFirstMultiplet} and 
\eqref{eq:ReviewSectionCFTPrimariesSecondMultiplet}). Our  bulk calculation only has information 
about the full conformal weight $\Delta=h+\bar{h}$ of the coset primaries, but the results 
correctly capture the dependence on the chiral conformal weights separately. This provides further 
evidence for the identification of states in \cite{Creutzig:2011fe}.

 Using the CFT, we have also obtained results for three-point functions involving fermionic
operators, and it would clearly be of interest to compare those with the corresponding bulk
quantities. This will require a slight generalization of our bulk techniques, in particular
in order to isolate the physical fermionic fields from the Vasiliev equations. This is 
currently under investigation.

 Of course, our approach of using a surrogate free-field CFT instead of the full-fledged 
Kazama-Suzuki model has severe limitations. It would be interesting to check whether  other
types of three-point functions (for instance, those involving three scalar fields) match 
between the bulk and the boundary theory. This would of course require analysing the matter 
Vasiliev equations beyond the linearised level. But it is unlikely that the free-field CFT 
can correctly capture those correlation functions, so any mismatch would be likely to be an
artifact of this. Even if one could reproduce all three-point
functions, the simple fact that the spectrum of the free theory is not the same as that of 
the $\mathbb CP^N$ model indicates that four-point functions will differ and matching those 
would require a more intricate boundary computation. Such checks would be essential in order to 
better establish the $\Ncal=2$ correspondence beyond the level of symmetries. 

 The way the correspondence is formulated at the moment, in order to go beyond the quantities
captured by the free CFT one would have to perform a computation at finite $N$ and $k$ and 
take the 't Hooft limit at the end. One might instead imagine a procedure by which one could 
obtain the nonlinear $\mathcal SW_\infty[\lambda]$ symmetry from the linear one directly in 
the 't Hooft limit, for instance by imposing a suitable constraint on the free CFT. This 
would probably provide a more efficient way to check the duality at large $N,k$. 

 Recently, an $\Ncal=1$ version of the higher-spin/minimal model correspondence was proposed
\cite{Creutzig:2012ar}. We expect the techniques used in this paper to transfer to that 
case with minor modifications, allowing the comparison of three-point functions in that 
model as well.

\subsection*{Acknowledgments}
We would like to thank  Matthias Gaberdiel, J\o rgen Rasmussen and Peter R\o nne for
discussions on higher-spin dualities. We would especially like to thank Niels Obers for many discussions 
and useful suggestions. 
Finally, we would like to thank the authors of \cite{Creutzigetal} for sharing a draft of their work
prior to publication. 
The work of KZ has been supported by FNU grant 272-08-0329 and by the Sapere Aude 
grant ``Strong Phases from First Principles''. This article is based on the first author's MSc project, 
submitted to the Niels Bohr Institute, University of Copenhagen, in November 2012. 

\appendix

\section[Brief Review of Vasiliev Theory on \TorPDF{\AdS_3}{AdS3}]{Brief Review of Vasiliev Theory on $\AdS_3$}\label{appendix:VasilievTheory}
In this appendix we intend to give a very brief, but self-contained, review of the full non-linear Prokushkin-Vasiliev theory as formulated in \cite{Prokushkin:1998bq,Prokushkin:1998vn}, for more details the reader is referred to these original papers and the review \cite{Vasiliev:1999ba}.

The full non-linear Vasiliev equations are formulated using an associative algebra $\mathcal A$ which is constructed using several auxiliary variables and a Moyal $\star$-product in the following way. Let $y_{\alpha}$ and $z_{\alpha}$ ($\alpha =1,2$) be two commuting bosonic twistor variables, where their spinor indices are raised and lowered as
  \begin{equation}
    y_\alpha  = y^\beta\epsilon_{\beta\alpha}, \qquad y^\alpha = \epsilon^{\alpha\beta}y_\beta,
  \end{equation}
where $\epsilon_{\alpha\beta}$ is the anti-symmetric tensor satisfying $\epsilon^{\alpha\beta}\epsilon_{\beta\gamma} = -\delta^\alpha_\gamma$. We will use the notation $uv=u_\alpha v^\alpha = -v_\alpha u^\alpha = -vu$ for contracted spinors. Beside these, we have two separate sets of Clifford elements $\psi_i$ ($i=1,2$) and $(k,\rho)$ satisfying the usual relations
  \begin{equation}\label{eq:AuxiliaryRelations1}
    \{\psi_i,\psi_j\} = 2\delta_{ij},\qquad \{k,\rho\} = 0,\qquad k^2=\rho^2=1.
  \end{equation}
All auxiliary variables commute with $\psi_{1,2}$, furthermore $\rho$ and $k$ commute and anti-commute with the twistor variables $y_\alpha$, $z_\alpha$, respectively
  \begin{equation}\label{eq:AuxiliaryRelations2}
    \qquad \{k,y_\alpha\} = 0,\qquad \{k,z_\alpha\} = 0,\qquad [\rho,y_\alpha]= 0,\qquad [\rho,z_\alpha] = 0.
  \end{equation}
A generic spacetime function mapping to this algebra has the following form
  \begin{equation}\label{eq:ExpansionOfFunctionsIntoAuxiliaryVariables}
    A(z,y;\psi_{12},k,\rho|x) = \sum_{B,C,D,E=0}^1\sum_{m,n=0}^{\infty}\frac 1{m!n!}\,A^{BCDE}_{\alpha_1\ldots\alpha_m\beta_1\ldots\beta_n}(x)\,k^B\rho^c\psi^D_1\psi^E_2 z^{\alpha_1}\dots z^{\alpha_m}y^{\beta_1}\dots y^{\beta_n}.
  \end{equation}
For our purposes, we will assume that the space-time functions $A^{BCDE}_{\alpha_1\ldots\alpha_m\,\beta_1\ldots\beta_n}(x)$ are symmetric in the spinor indices. Furthermore the Grassmann parity of the coefficients $A^{BCDE}_{\alpha_1\ldots\alpha_m\,\beta_1\ldots\beta_n}(x)$ is equal to the number of spinor indices mod 2 and they are defined to commute with all the generating elements $y_\alpha$, $z_\alpha$, $k$, $\rho$ and $\psi_{1,2}$. Thus commutators of functions of the form \eqref{eq:ExpansionOfFunctionsIntoAuxiliaryVariables}, will automatically turn into supercommutators of polynomials of $y_\alpha$, $z_\alpha$, $k$, $\rho$ and $\psi_{1,2}$.

In order to formulate the theory, we also need the $\star$-product defined on functions of $y$ and $z$ given by
  \begin{equation}
    (f\star g)(z,y) = \frac 1{(2\pi)^2}\int\ud^2u\,\ud^2v\,\exp(iu_\alpha v^\alpha)\,f(z+u,y+u)\, g(z-v,y+v).
  \end{equation}
This product turns out to be associative and have a regularity property, the product of two polynomials will also be a polynomial in $y$ and $z$. 
Defining the $\star$-commutator $[V,W]_\star = V\star W- W\star V$, we have the important commutators
  \begin{equation}
    [y_\alpha,y_\beta]_\star = -[z_\alpha,z_\beta]_\star = 2i\epsilon_{\alpha\beta}, \qquad [y_\alpha,z_\beta]_\star =0.
  \end{equation}
It turns out that the basic variables $y_\alpha$ and $z_\alpha$ behave as derivatives, in particular for a very general class of functions \cite{Prokushkin:1998bq} we have $[y_\alpha, f]_\star = 2i\diffp f{y^\alpha}$ and $[z_{\alpha},f]_\star = -2i\diffp f{z^\alpha}$.
Note that the star product only operates on the twistor components, but the order of all auxiliary variables is important due to the relations \eqref{eq:AuxiliaryRelations1} and \eqref{eq:AuxiliaryRelations2}.

Vasiliev theory is formulated in terms of three generating functions depending on spacetime coordinates and the auxiliary variables
    \begin{gather}
      \begin{split}
	W &= W_\mu(z,y;\psi_{1,2},k,\rho|x)\,\ud x^\mu,\\
	B &= B(z,y;\psi_{1,2},k,\rho|x),\\
	S_\alpha &= S_\alpha(z,y;\psi_{1,2},k,\rho|x).
      \end{split}
    \end{gather}
The spacetime 1-form $W$ is the generating function of the higher-spin fields, the 0-form $B$ is the generating function of the massive matter fields while $S_\alpha$ describes pure gauge degrees of freedom and is necessary for consistent internal symmetries. The full set of non-linear Vasiliev equations is then given by
    \begin{gather}\label{eq:FullNonLinearVasilievEquationOfMotion}
      \begin{aligned}
	\ud W - W\star\wedge W &=0,\\
	\ud B + [B,W]_\star &= 0, \\
	\ud S_\alpha + [S_\alpha,W]_\star &=0,\\
	S_\alpha\star S^\alpha + 2i(1+B\star K) &= 0,\\
	[S_\alpha, B]_\star &=0,
      \end{aligned}
    \end{gather}
where $K= k e^{zy}$ is the Kleinian. The last two constraints guarantee that local Lorentz invariance remains unbroken to all orders of interaction. It turns out that due to the involutive automorphism $\rho\rightarrow -\rho$, $S_\alpha\rightarrow -S_\alpha$ one can truncate the system such that $W$ and $B$ become $\rho$-independent, while $S_\alpha(z,y;\psi_{1,2},k,\rho|x) = \rho\,s_\alpha(z,y;\psi_{1,2},k|x)$. This is the system studied in this paper and in \cite{Prokushkin:1998bq,Prokushkin:1998vn}.
One can readily check that these equations are invariant under the following set of $\rho$-independent local higher gauge transformations, parametrized by $\epsilon = \epsilon(z,y;\psi_{1,2},k|x)$
    \begin{gather}\label{eq:InfinitesimalGaugeTransformationsFullVasilievTheory}
      \begin{aligned}
	\delta W &= \ud\epsilon +[\epsilon,W]_\star,\\
	\delta B &= [\epsilon, B]_\star,\\
	\delta S_\alpha &= [\epsilon, S_\alpha]_\star.
      \end{aligned}
    \end{gather}
Note that the equations of motion and gauge transformations for the higher-spin fields look very similar to usual Chern-Simons theory. As mentioned earlier, the commutators in \eqref{eq:FullNonLinearVasilievEquationOfMotion} and \eqref{eq:InfinitesimalGaugeTransformationsFullVasilievTheory} are actually supercommutators of polynomials of the generating elements, $y_\alpha$, $z_\alpha$, $k$ and $\psi_{1,2}$.

\subsection{Vacuum solutions and linearized dynamics}
Now we consider vacuum solutions of the Vasiliev equations \eqref{eq:FullNonLinearVasilievEquationOfMotion} in which the matter fields take a constant value
  \begin{equation}\label{eq:ConstantVacuumSolutionForMatterFields}
    B^{(0)} = \nu = \text{constant}.
  \end{equation}
With this ansatz the second and the last equations of \eqref{eq:FullNonLinearVasilievEquationOfMotion} are trivially satisfied, while the vacuum solutions of $W$ and $S_\alpha$ have to satisfy the three remaining ones
  \begin{gather}\label{eq:VasilievVacuumEquations}
    \begin{aligned}
      \ud W^{(0)} - W^{(0)}\star\wedge W^{(0)} &=0,\\
      \ud S_\alpha^{(0)} + [S_\alpha^{(0)},W^{(0)}] &=0,\\
      S_\alpha^{(0)}\star S^{(0)\alpha}  + 2i(1+\nu K) &= 0.
    \end{aligned}
  \end{gather}
In \cite{Prokushkin:1998bq} three different solutions to the third equation are given but they are all in the same gauge equivalence class, the simplest is
  \[ S^{(0)}_\alpha = \rho\,\tilde z_\alpha, \qquad \text{where}\qquad \tilde z_\alpha = z_\alpha + \nu(z_\alpha + y_\alpha)\int_0^1\ud t\, t\,e^{it\,zy}\,k\;. \]
Since $\ud S_\alpha^{(0)}=0$, the second equation of \eqref{eq:VasilievVacuumEquations} reduces to $[S^{(0)}_\alpha,W^{(0)}] = 0$. In order to solve this constraint, one can show that the following element
  \begin{gather}
    \tilde y_\alpha = y_\alpha + \nu(z_\alpha + y_\alpha)\int_0^1\ud t\, (t-1)\, e^{it\,yz}\, k,
  \end{gather}
satisfies the commutation relations
  \begin{equation}\label{eq:DeformedOscillatorCommutatorAppendix}
    [\tilde y_\alpha,\tilde y_\beta]_\star = 2i\epsilon_{\alpha\beta}(1 + \nu k), \qquad \{\tilde y,k\} =0,
  \end{equation}
and finally $[\tilde y_\alpha, S^{(0)}_\beta] = 0$. Now the constraint $[S^{(0)}_\alpha,W^{(0)}]=0$ is solved if $W^{(0)}$ depends only on $\psi_{1,2}$, $k$ and $\tilde y_\alpha$, since they all commute with $S^{(0)}_\alpha$.

Note the remarkable feature of \eqref{eq:DeformedOscillatorCommutatorAppendix}, the vacuum constant $\nu$ is deforming the oscillators $y_\alpha$ into the so-called deformed oscillators $\tilde y_\alpha$. This means that $\nu$ is parametrizing a continuous class of vacuum solutions \eqref{eq:ConstantVacuumSolutionForMatterFields}, in which the symmetry algebra is continuously deforming. As we see in section \ref{Subsection:ScalarsOnAdS3}, $\nu$ also fixes the masses of the matter fields. We will call the associative algebra generated by $\tilde y_\alpha$, $k$ and $\psi_{1,2}$ for $\mathcal A_S$.

Next we will consider linearized fluctuations of the matter fields around this vacuum, propagating on the higher-spin background $W^{(0)}$
  \begin{equation}\label{eq:MatterFieldsLinearFluctuationsAroundVacuum}
    B(z,y;\psi_{1,2},k) = \nu + \mathcal C(z,y;\psi_{1,2},k).
  \end{equation}
In this paper we will neglect all fluctuations around $W^{(0)}$ and $S_\alpha^{(0)}$, thus we do not consider higher order effects like backreaction of the matter on the higher-spin fields and interactions among the matter fields. See \cite{Prokushkin:1998bq} for more about these issues. Inserting \eqref{eq:MatterFieldsLinearFluctuationsAroundVacuum} into \eqref{eq:FullNonLinearVasilievEquationOfMotion} we get two non-trivial equations
  \begin{gather}\label{eq:LinearizedEQNAppendix}
    \begin{split}
      \ud\mathcal C + [\mathcal C, W^{(0)}]_\star &= 0,\\
      [S_\alpha^{(0)}, \mathcal C]_\star &= 0.
    \end{split}
  \end{gather}
The second equation is solved by demanding that $\mathcal C$ is a spacetime function mapping into the algebra $\mathcal A_S$. In other words, we have found that $\mathcal C=\mathcal C(\tilde y;k,\psi_{1,2}|x)$ and $W^{(0)} = W^{(0)}(\tilde y;k,\psi_{1,2}|x)$. We can now get rid of the $\psi_{1,2}$ Clifford elements and find the equations of motion of the physical fields. For this we need to define the projection operators
  \begin{equation}
    \mathcal P_\pm =\frac{1\pm\psi_1}2,
  \end{equation}
with the following properties
  \begin{equation}
    \mathcal P_\pm\mathcal P_\mp = 0,\qquad \mathcal P_\pm^2 = \mathcal P_\pm.
  \end{equation}
The usual gauge fields  $A$ and $\bar A$, known from $\AdS_3$ gravity, are extracted as
  \[
    W^{(0)} = -\mathcal P_+ A - \mathcal P_-\bar A.
  \]
Inserting this into the equations of motion for $W^{(0)}$ \eqref{eq:VasilievVacuumEquations}, we find the Chern-Simons flatness conditions
  \begin{equation}
    \ud A + A\star\wedge A = 0,\qquad \ud\bar A + \bar A\star\wedge\bar A = 0.
  \end{equation}
The matter fields can be decomposed as
  \begin{equation}
    \mathcal C(\tilde y;k,\psi_{1,2}|x) = \mathcal C_{aux}(\tilde y;k,\psi_1|x) + \mathcal C_{dyn}(\tilde y;k,\psi_1|x)\psi_2.
  \end{equation}
It turns out that $C_{aux}$ does not describe any propagating degrees of freedom and can consistently be put to zero and the dynamical part $C_{dyn}$ can be decomposed as
  \begin{equation}
    \mathcal C(\tilde y;k,\psi_1|x) = C(\tilde y;k|x)\,\mathcal P_+\,\psi_2 + \tilde C(\tilde y;k|x)\,\mathcal P_-\,\psi_2.
  \end{equation}
The equations \eqref{eq:LinearizedEQNAppendix} finally reduce to
  \begin{gather}\label{eq:MatterEQMAppedix}
    \begin{aligned}
      \ud C+A\star C - C\star\bar A &=0,\\
      \ud\tilde C + \bar A\star\tilde C-\tilde C\star A &= 0.
    \end{aligned}
  \end{gather}
The associative algebra generated by $\tilde y^\alpha$ and $k$ modulo the relations \eqref{eq:DeformedOscillatorCommutatorAppendix} is known as $Aq(2,\nu)$ \cite{Vasiliev:1989qh}. The physical fields in this algebra are expanded as
  \begin{gather}
    \begin{aligned}
      C(\tilde y;k|x) &= \sum_{B=0}^1\sum_{n=0}^\infty \frac 1{n!} C^B_{\alpha_1\dots\alpha_n}(x)\,k^B\,\tilde y^{\alpha_1}\star\dots\star\tilde y^{\alpha_n},\\
      A(\tilde y;k|x) &= \sum_{B=0}^1\sum_{n=0}^\infty \frac 1{n!} A^B_{\alpha_1\dots\alpha_n}(x)\,k^B\,\tilde y^{\alpha_1}\star\dots\star\tilde y^{\alpha_n},
    \end{aligned}
  \end{gather}
and similarly for $\tilde C$ and $\bar A$. The element $k$ doubles the spectrum, this is needed in order to have $\mathcal N=2$ supersymmetry. The lowest components of $C$ with no spinor index correspond to two scalars, while the ones with one spinor index correspond to two fermions and similarly for $\tilde C$. The functions $C^B_{\alpha_1,\dots, \alpha_n}$, for $n>1$, are auxiliary fields and can all be written as sums of derivatives of the physical fields, using the equations of motion \eqref{eq:MatterEQMAppedix}.

The algebra $Aq(2,\nu)$ contains the subalgebra of even elements $C(\tilde y;k|x) = C(-\tilde y;k|x)$, which can be decomposed as $Aq^E(2,\nu)\oplus Aq^E(2,-\nu)$ \cite{Vasiliev:1989qh} by the projection operator $\Pi_\pm = \frac{1\pm k}2$. Thus one obtains a non-supersymmetric truncation by restricting to even polynomials of $\tilde y_\alpha$ and projecting $k=\pm 1$, this was recently used in \cite{Ammon:2011ua}. There also exists a $\mathcal N=1$ truncation \cite{Prokushkin:1998bq,Prokushkin:1998vn}.

\section[The \TorPDF{\mathcal{SB}[\mu]}{SB[mu]} and \TorPDF{\shs[\lambda]}{shs[lambda]} Algebras]{The $\mathcal{SB}[\mu]$ and $\shs[\lambda]$ Algebras} \label{appendix:StructureConstantsOfHigherSpinAlgebra}
This appendix contains information and definitions of functions related to the algebras $\mathcal{SB}[\mu]$ and $\shs[\lambda]$, together with a few useful properties. These algebras were briefly defined in section \ref{eq:ModifiedVasilievFormalism}. For the structure constants of the infinite dimensional associative algebra $\mathcal{SB}[\mu]$, we will use the following notation
  \begin{gather}\label{eq:StructureConstantsAssociativeHigherSpinAlgebra}
    \begin{aligned}
      L_m^{(s)} &\star L_n^{(t)}	=	\sumCircles_{u=1}^{s+t-1} g_u^{st}(m,n;\lambda)\:L_{m+n}^{(s+t-u)},\\
      G_p^{(s)} &\star G_q^{(t)}	=	\sumCircles_{u=1}^{s+t-1}\tilde g_u^{st}(p,q;\lambda)\:L_{p+q}^{(s+t-u)},
    \end{aligned}
	\hspace{8mm}
    \begin{aligned}
      L_m^{(s)} &\star G_q^{(t)}	=	\sumCircles_{u=1}^{s+t-1} h_u^{st}(m,q;\lambda)\:G_{m+q}^{(s+t-u)},\\
      G_p^{(s)} &\star L_n^{(t)}	=	\sumCircles_{u=1}^{s+t-1} \tilde h_u^{st}(p,n;\lambda)\:G_{p+n}^{(s+t-u)},
    \end{aligned}
  \end{gather}
while for the infinite dimensional Lie superalgebra $\shs[\lambda]$ we use the notation
  \begin{gather}
    \begin{aligned}
        &\left[L_m^{(s)}, L_n^{(t)}\right]	=	\sumCircles_{u=1}^{s+t-1}\hat g_u^{st}(m,n;\lambda)\:L_{m+n}^{(s+t-u)},\\
	&\left\{G_p^{(s)},  G_q^{(t)}\right\}	=	\sumCircles_{u=1}^{s+t-1}\hat{\tilde g}_u^{st}(p,q;\lambda)\:L_{p+q}^{(s+t-u)},\\
    \end{aligned}
	\hspace{8mm}
    \begin{aligned}
       &\left[L_m^{(s)},  G_q^{(t)}\right]	=	\sumCircles_{u=1}^{s+t-1}\hat h_u^{st}(m,q;\lambda)\:G_{m+q}^{(s+t-u)},\\
       &\left[G_p^{(s)},  L_n^{(t)}\right]	=	\sumCircles_{u=1}^{s+t-1}\hat{\tilde h}_u^{st}(p,n;\lambda)\:G_{p+n}^{(s+t-u)},\\
    \end{aligned}
  \end{gather}
where the notation $\sumCircles$ stands for sum over half-integer steps.
If one does not put any constraints on the modes, this then corresponds to  the linear $sw_{\infty}[\lambda]$ algebra. If one restricts to the wedge subalgebra, one can show that it is safe to restrict the sums to $1\leq u\leq \text{Min}(2s-1,2t-1)$ since the structure constants for higher $u$ vanish (this is not the case for modes outside the wedge).

\subsection{Useful properties of structure constants}
For reference, we will in this section list a few properties of some of the $\mathcal{SB}[\mu]$ and $\shs[\lambda]$ structure constants which are quite useful for our calculations.
  \begin{gather}
     \begin{split}
	g^{st}_u(m,n;\lambda) = \begin{cases}
	                          (-1)^{\floor u + 1}\: g^{ts}_u(n,m;\lambda) &\begin{cases}
										u\in\mathbb Z,  &\big(s,t\in\mathbb Z \hspace{11mm}\text{or}\quad s+t\in\mathbb Z+\frac 12\big)\\
										u\in\mathbb Z+\frac 12,  &\big(s,t\in\mathbb Z+\frac 12 \quad\text{or}\quad s+t\in\mathbb Z+\frac 12\big)
									     \end{cases}\\
	                          (-1)^{\floor u}\: g^{ts}_u(n,m;\lambda) &\begin{cases}
										u\in\mathbb Z,  &s,t\in\mathbb Z+\frac 12\\
										u\in\mathbb Z+\frac 12,  &s,t\in\mathbb Z
									     \end{cases}
	                        \end{cases}
     \end{split}
  \end{gather}

  \begin{gather}
    \begin{aligned}
      g^{st}_1(m,n;\lambda) &= \begin{cases}
				1 \qquad \big(s,t\in\mathbb Z\quad \text{or}\quad s+t\in\mathbb Z+\frac 12\big)\\
				0 \qquad s,t\in\mathbb Z+\frac 12
			      \end{cases}\\ 
      g^{st}_{\frac 32}(m,n;\lambda) &= \begin{cases}
				m/2\;\;\text{or}\;\; n/2 \qquad &\big(s=1,\; t\in\mathbb Z\big)\quad\text{or}\quad\big(s\in\mathbb Z,\; t=1\big)\\
				0 \qquad &s,t\in\mathbb Z\quad \text{and}\quad s,t\neq 1\\
				g^{st}_{\frac 32}(0,0;\lambda)	& \big(s,t\in\mathbb Z+\frac 12\big)\quad\text{or}\quad \big(s+t\in\mathbb Z+\frac 12\big)
			      \end{cases}
    \end{aligned}
  \end{gather}
  \begin{gather}\label{eq:StructureConstantPropertyForOPECalculation}
    \begin{aligned}
     \hat g^{2s}_u(1,m;\lambda) &= \begin{cases}
				  \floor s - 1 - m,  &u=2\\
				  0,		     &u=1,\frac 32,\frac 52,3
                               \end{cases},\\
      \hat h^{2s}_u(1,r;\lambda) &= \begin{cases}
				  \ceil s - \frac 32 - r,  &u=2\\
				  0,		     &u=1,\frac 32,\frac 52,3
                               \end{cases}.
    \end{aligned}
  \end{gather}

\subsection[Structure Constants of \TorPDF{\mathcal{SB}[\mu]}{SB[mu]}]{Structure Constants of $\mathcal{SB}[\mu]$}
In this section we will list explicit formulas for the structure constants of the infinite dimensional associative algebra, $\mathcal{SB}[\mu]$. See the appendix of \cite{Hanaki:2012yf} for a sketch of how these are derived from the results of \cite{Bergshoeff:1991dz}.
The $L\star L$ structure constant is given as\footnote{The various functions appearing in this and the following expressions 
are listed at the end.}
\begin{multline}\label{eq:StructureConstantg}
  g_u^{st}(m,n;\lambda) = \sum_i F_{st}^u\left[\hh{u}{\hhti{s+t}{}}i+
	    \hti{s}\hhti{u}{\hhti{s+t}{}};\lambda\right]\\
    \times\left(m-\floor s + 1\right)_{\ceil{i,u,s,t}_1} \left(n - \floor t +1\right)_{\floor u - 1 +\hhti{s}{}
    \hhti{t}{}-\hhti{u}{}\hhti{s+t}{}-\ceil{i,u,s,t}_1},
\end{multline}
where the range of the sum is
\[
  0\leq i \leq \hh{u}{\hti{s+t}}\left(\floor u -1\right) + \hti{u}\hhti{s+t}{} - \hti{s}\hhti{u}{\hhti{s+t}{}}\hhti{u}{}.
\]
Similarly we have for the $G\star G$ product
\begin{multline}\label{eq:StructureConstantgTilde}
  \tilde g_u^{st}(p,q;\lambda) = -\hh{s}{}\hh{t}{}\sum_i (-1)^{\left[i+\hti{s}\right]\hhti{u}{\hti{s+t}}}\\
   \times F_{st}^u\left[\hh{u}{\hhti{s+t}{}}i+
	    \hhti{s}{}\hhti{u}{\hhti{s+t}{}};\lambda\right]\\
    \times\left(p-\ceil s + \tfrac 32\right)_{\ceil{i,u,s,t}_2} \left(q - \ceil t +\tfrac 32\right)_{\floor u - \hhti{s}{} -\hti{s+t}
    \hti{s}-\hhti{s+t}{}\hhti{u}{}-\ceil{i,u,s,t}_2},
\end{multline}
where,

\begin{multline}
 0 \leq i \leq \hh{u}{\hti{s+t}}\left(u -1\right) - \left[\hhti{s}{}+\hti{s+t}\hti{s}\right]\hhti{u}{\hhti{s+t}{}}\\
					\times\left(\hhti{u}{}+\tfrac 12\hti{s+t}\right).
\end{multline}
And for $L\star G$
\begin{multline}\label{eq:StructureConstanth}
  h_u^{st}(m,q;\lambda) = \hhh{u}{\hti{s}}{(-1)^{\hti{t}}}\sum_i F_{st}^u\Big[\hh{u}{\hhti{s}{\hti{t}}}i+
	    \hti{s}\\\times\hhti{u}{\hhti{t}{}};\lambda\Big]\\
    \times\left(m-\floor s + 1\right)_{\ceil{i,u,s,t}_3} \left(q - \ceil t +\tfrac 32\right)_{\floor u - \hhti{t}{} -\hti{t}
    \hti{s}-\hhti{s}{\hti{u}}\hhti{u}{}-\ceil{i,u,s,t}_3},
\end{multline}
where,

\begin{multline}
  0\leq i \leq \hh{u}{\hhti{s}{\hhti{t}{}}}\left(u -1\right) - \hti{s}\hhti{u}{\hhti{t}{}}\hhti{u}{}\\ -\tfrac 12 \hhti{s}{\hhti{t}{}}\hti{u}.
\end{multline}
And finally for the $G\star L$ product
\begin{multline}\label{eq:StructureConstanthTilde}
  \tilde h_u^{st}(p,n;\lambda) = \hhh{u}{\hti{t}}{(-1)^{\hti{s}}}\sum_i (-1)^{\left[i+\hti{s}\right]\hhti{u}{\hhti{t}{\hhti{s}{}}}}\\
     F_{st}^u\left[\hh{u}{\hhti{t}{\hti{s}}}i+
	    \hhti{s}{}\hhti{u}{\hti{t}};\lambda\right]\\
    \times\left(p-\ceil s + \tfrac 32\right)_{\ceil{i,u,s,t}_4} \left(n - \floor t + 1\right)_{\floor u  -\hhti{s}{}
	-\hti{s}\hti{t}-\hhti{t}{\hti{s}}\hhti{u}{}-\ceil{i,u,s,t}_4},
\end{multline}
where,
\begin{multline}
  0\leq i \leq \hh{u}{\hhti{t}{\hhti{s}{}}}\left(u -1\right) - \hhti{s}{}\hhti{u}{\hti{t}}\hhti{u}{}\\ -\tfrac 12 \hhti{t}{\hhti{s}{}}\hhti{u}{\hhti{s}{}\hti{t}}.
\end{multline}
The functions used in the above structure constants are
  \begin{gather}
    \begin{align}
      \ceil{i,u,s} 		&= \ceil[\Bigg]{\h{u}\frac{\big[i+\hhti{u}{}\hti{s}\big]}{2}}\nonumber\\
      \ceil{i,u,s,t}_1 	&= \ceil[\big]{i,u+\tfrac 12\hhti{s+t}{},s}\\
      \ceil{i,u,s,t}_2 	&= \ceil[\Big]{i,u+\tfrac 12\hhti{s+t}{},s+\tfrac 12\hhti{s+t}{}+\tfrac 12\hti{s+t}\left\{\hhti{s}{}+\hti{s}\hhti{u}{}\right\}}\nonumber\\				
      \ceil{i,u,s,t}_3 	&= \ceil[\Big]{i,u+\tfrac 12\hhti{s}{\hti{t}},s+\tfrac 12\hti{t}\hti{s}\hti{u}}\nonumber\\
      \ceil{i,u,s,t}_4 	&= \ceil[\Big]{i,u+\tfrac 12\hhti{t}{\hti{s}},s+\tfrac 12}\nonumber
    \end{align}
  \end{gather}
where 
  \begin{gather}
    \begin{aligned}
      \h u		&= \ceil[\big]{u-\floor u + 1}\;,\\
      \hti u	&= \ceil[\big]{u-\floor u }\;.
    \end{aligned}
  \end{gather}
We also use the definitions
\begin{align}
  |n|_2	&= n-2\floor{n/2},\\
  (a)_n	&= a(a+1)(a+2)\dots (a+n-1),\qquad(a)_0 = 1,\\
  \begin{bmatrix}
      a \\ b
  \end{bmatrix}
	&= \frac{\floor{a}!}{\floor{b}!\floor{a-b}!}\;,
\end{align}
where $(a)_n$ is the ascending Pochhammer symbol. Finally, the various intermediate coefficients are defined as
\begin{align}
  F^u_{st}(\lambda) &= (-1)^{\floor{s+t-u-1}}\frac{(2s+2t-2u-2)!}{(2s+2t-\floor{u}-3)!}\sum_{i=0}^{2s-2}\sum_{j=0}^{2t-2}\delta(i+j-2s-2t+2u+2)\\
		    &\quad\times A^i(s,\tfrac 12-\lambda)A^j(t,\lambda)(-1)^{2s+2i(s+t-u)},\nonumber\\
  A^i(s,\lambda)    &= (-1)^{\floor{s}+1+2s(i+1)}\begin{bmatrix}
						  s-1 \\ i/2
                                                 \end{bmatrix}
  \frac{\left([(i+1)/2]+2\lambda)\right)_{\floor{s-1/2}-\floor{(i+1)/2}}}{\left(\floor{s+i/2}\right)_{2s-1-\floor{s+i/2}}},\label{eq:DefinitionOfTheAFunction}
\end{align}
and 
\begin{align}
  F^u_{st}(i,\lambda) &= F^u_{st}(\lambda)(-1)^{\floor{i/2}+2i(s+u)}\begin{bmatrix}
								    u-1 \\ i/2
								  \end{bmatrix}
    \left(\floor{2s-u}\right)_{\floor{u-1-i/2}+|2u|_2|2u-2-i|_2}\\
  &\times \left(\floor{2t-u}\right)_{\floor{i/2}+|2u|_2|i|_2}\nonumber\;.
\end{align}

\subsection[Structure Constants of \TorPDF{\shs[\lambda]}{shs[lambda]}]{Structure Constants of $\shs[\lambda]$}
These structure constants are directly given by the formulas for the $\mathcal{SB}[\mu]$ structure constants, but the constants $F^u_{st}(\lambda)$ have to be replaced by
  \begin{equation}
    f^u_{st}(\lambda) = F^u_{st}(\lambda) + (-1)^{\floor{-u}+4(s+u)(t+u)}\, F^u_{st}(\frac 12-\lambda).
  \end{equation}

\bibliographystyle{utcaps}
\bibliography{refs}

\end{document}